\documentstyle[12pt]{article}  

\begin{document}

\newtheorem{thm}{Theorem}[section]
\newtheorem{lemma}[thm]{Lemma}
\newtheorem{prop}[thm]{Proposition}
\newtheorem{defin}[thm]{Definition}
\newtheorem{cor}[thm]{Corollary}

 %
 %
 %
 %

\newcommand{\lap}{\bigtriangleup}
\def\be{\begin{equation}}
\def\ee{\end{equation}}
\def\bea{\begin{eqnarray}}
\def\eea{\end{eqnarray}}
\def\beas{\begin{eqnarray*}}
\def\eeas{\end{eqnarray*}}

\def\r{\rho}
\def\s{\sigma}
\def\e{\epsilon}
\def\z{\zeta}
\def\g{\gamma}
\def\x{\xi}
\def\O{\Omega}

\def\d#1{\partial_{#1}}
\def\dt{\partial_t}
\def\dx{\partial_x}
\def\dv{ \partial_v }
\def\dz{ \partial_\z }
\def\R{\mathrm{ I\kern-.1567em R}}
\def\N{{\rm I\kern-.1567em N}}
\def\Z{{\sf Z\kern-.3567em Z}}
\def\rdd{\dot{\mathrm{ I\kern-.1567em R}}^{\raisebox{-2.5pt}
{$\scriptstyle 3$}}}

\def\open#1{\setbox0=\hbox{$#1$}
\baselineskip = 0pt
\vbox{\hbox{\hspace*{0.5 \wd0}\tiny $\circ$}\hbox{$#1$}} 
\baselineskip = 12pt\!}

\def\supp{\mbox{\rm supp}} 
\def\div{\mbox{\rm div}}
\def\n#1{\vert #1 \vert}
\def\nn#1{{\Vert #1 \Vert}}
\def\prf{\noindent
         {\bf Proof:\ }}
\def\prfe{\hspace*{\fill} $\bullet$ 

\smallskip \noindent}

\title{Stationary and static stellar dynamic models with axial symmetry} 
\author{Gerhard Rein\\
        Department of Mathematics, Indiana University\\
        Bloomington, IN 47\,405, U.S.A.\\
        On leave from:\\
        Mathematisches Institut der Universit\"at M\"unchen\\
        Theresienstr.\ 39, 80333 M\"unchen, Germany}  
\date{}
\maketitle
\begin{abstract}
We study the existence of stationary solutions of the Vlasov-Poisson system
with finite radius and finite mass in the stellar dynamics case. So far, 
the existence of such solutions is only known under the assumption of
spherical symmetry. Using the implicit function theorem we show
that certain stationary, spherically symmetric solutions can be embedded 
in one parameter families of stationary, axially symmetric solutions with 
finite radius and finite mass. In general, these new steady states
have non-vanishing average velocity field, but they can also be constructed 
such that their velocity field does vanish, in which case they are called 
static. 
\end{abstract}
\section{Introduction}
\setcounter{equation}{0}
Large stellar systems such as galaxies or globular clusters
are often described by a density function  $f = f(t,x,v)\geq 0$ on phase 
space;
$t \in \R$ denotes time and $x, v \in \R^3$ denote position and velocity
respectively. Under the assumption that the mass points in the ensemble, 
i.~e., the stars, interact only by the gravitational potential which they
create collectively and that in particular collisions are negligible,
the time evolution of the ensemble is described by the following nonlinear 
system of partial differential equations, known as the Vlasov-Poisson system:
\[
\dt f + v\cdot \nabla_x f - \nabla_x U \cdot \nabla_v f =0,
\]
\[ 
\lap U = 4\pi \r,\ \r (t,x) = \int f(t,x,v)\,dv .
\]
Here $U = U(t,x)$ denotes the gravitational potential of the ensemble
and $\r = \r (t,x)$ denotes its spatial mass density. 
For simplicity we assume that all particles have the same 
mass, equal to unity, and set the gravitational constant equal to
unity as well. 

In the present investigation we are interested in the existence
and properties of solutions of this system, which
are independent of time.
Such solutions are usually called stationary. If they have the additional
property that  their average velocity field vanishes, i.~e.,
$\int v f(x,v)\, dv / \r (x) =0,\ x \in \R^3$, we shall call them static. 
If $U$ is independent of time, the energy 
\be \label{edef}
E(x,v) := \frac{1}{2} v^2 + U(x)
\ee
of a particle with coordinates $(x,v) \in \R^6$
is constant along solutions of the characteristic system
\[
\dot x = v,\ \dot v = - \nabla U (x)
\]
of the Vlasov equation. Therefore, the ansatz
\[
f(x,v) = \Phi (E)
\]
automatically satisfies the Vlasov equation and reduces the problem
of finding a stationary solution of the Vlasov-Poisson system to solving
the semilinear elliptic problem
\be \label{nopo}
\lap U = 4 \pi h_\Phi (U)
\ee
where $h_\Phi$ is obtained by inserting the ansatz for $f$ into the definition
of $\r$. If other invariants of the characteristic flow are 
known---such as the modulus of the angular momentum $\n{x \times v}$ in
case of spherical symmetry---then $\Phi$ can depend on these additional
invariants as well and the right hand side of (\ref{nopo})
can become explicitly $x$-dependent. The main difficulty 
with this approach is to show that 
a solution of (\ref{nopo})---once its existence is established---leads
to a stationary model with physically reasonable properties, in particular,
with finite mass and finite radius, i.~e., $\r$ is compactly supported.
In \cite{BFH} this program was carried out under the assumption of
spherical symmetry, where it can also be shown that the distribution function
$f$ must be of the form $f(x,v) = \Phi (E, \n{x \times v})$ for some $\Phi$.
Spherically symmetric stationary solutions are automatically static.
To the best of our knowledge, the existence
of static or even stationary solutions of the Vlasov-Poisson
system in the stellar dynamics case with finite radius and finite mass
and without spherical symmetry is open. 
This is interesting because for a selfgravitating fluid it
is known that every static solution must be spherically symmetric,
cf.~\cite{L2}. 

In the present paper we show that this is not so for a selfgravitating
ensemble as described by the Vlasov-Poisson system. In fact, we will show
that every static solution $(f_0, \r_0, U_0)$ in a certain subclass of the 
spherically symmetric ones obtained in \cite{BFH} is embedded in continuous
one-parameter families $(f^\g,\r^\g,U^\g)$ of stationary solutions with axial 
symmetry which coincide with $(f_0,\r_0,U_0)$ for $\g = 0$, have finite radius 
and finite mass, and are not spherically symmetric for $\g \neq 0$. 
Families of static as well as families of stationary but not static such 
solutions are obtained for the same spherically symmetric steady state.
For the precise statement of our result we refer to the next section.
The basic idea of the proof is the following: Assuming that
$U$ is axially symmetric, i.~e., $U(Ax) = U(x)$ for every $x \in \R^3$
and every rotation $A$ around, say, the $x_3$-axis, the quantity
\be \label{pdef}
P(x,v) := x_1 v_2 - x_2 v_1 ,
\ee
that is the $x_3$-component of the angular momentum of
the particle with coordinates $(x,v) \in \R^6$, is conserved
along characteristics. We make the ansatz 
\[
f(x,v) = \Phi (E, \gamma P)
\]
with $\Phi$ such that $\gamma = 0$ leads to one of the spherically
symmetric solutions with finite radius and finite mass obtained in \cite{BFH}.
Then we transform the problem (\ref{nopo}) to the problem of
finding zeros of an operator $T(\gamma,\cdot)$ where for $\g = 0$ we know 
a zero, namely $U_0$, and we can try to prove our result
by applying the implicit function theorem. The central idea which
makes this approach work is to look for $U^\g$ as a deformation
of $U_0$, i.~e., $U^\g (x) = U_0 (g(x))$ for some diffeomorphism
$g$ on $\R^3$, and to formulate the problem of finding zeros of
$T$ over the space of such deformations instead of
the space of potentials. Whereas the original problem (\ref{nopo})
had to be solved on $\R^3$, it turns out that one needs to know
the deformation only on a compact neighbourhood of
the support of the original solution
$(f_0,\r_0,U_0)$, and this provides useful compactness
properties. 
In particular, this deformation approach is essential in proving that
the derivative of $T$ at $U_0$ is an isomorphism.
Finite radius and finite mass of the resulting
stationary solutions are then  immediate consequences of the
corresponding properties of $(f_0,\r_0,U_0)$.

The approach which we explained above has been used by {\sc Lichtenstein}
for proving the existence of slowly rotating Newtonian stars, as
described by selfgravitating fluid balls, cf.~\cite{L1,L2}.
A translation of {\sc Lichtenstein's} approach into
modern mathematical language and the framework of the implicit function
theorem is due to {\sc Heilig}, cf.~\cite{H}, and the present paper
owes much to that investigation. Our approach is analogous to
\cite{H} but the actual proofs are different, so that we decided
to give a self-contained presentation of the arguments for the present case 
of the Vlasov-Poisson system. Our paper proceeds as follows:
In the next section we formulate our result and the general framework for its proof. 
In particular, we
define the Banach spaces which serve
as domain and range for the operator $T(\gamma,\cdot)$, introduce the deformation 
mappings and  show how 
our result is obtained from the implicit
function theorem. The continuous Fr\'{e}chet-differentiability
of $T$ with respect to the second argument and the fact that at zero this
derivative is an isomorphism are then established
in Sections 3 and 4 respectively.
     
We conclude this introduction with some references to the now
quite extensive literature on the Vlasov-Poisson system. 
Global existence of classical solutions has been established in 
\cite{Pf}, cf.\ also \cite{Ho,LP,RR,Sch} and the review article
\cite{R3}. For the plasma physics case, where the sign of the source 
term in the Poisson equation is reversed, the existence of stationary
solutions, say, on bounded domains or with a fixed ion background
or external force field, is much easier to obtain, cf.\
for example \cite{R1}. Moreover,  there are now several results
on the stability properties of such stationary plasmas, cf.\ for example
\cite{GS,R2}. The stability question for the stellar dynamics
case is much harder, and preliminary results can be found in
\cite{BMR,Wo}, cf.~also \cite{R4}. 
Coming back to the topic of the present paper
we mention that in \cite{BBDP} families of stationary solutions
of the stellar dynamic Vlasov-Poisson system with axial symmetry
were obtained, but these models have infinite mass and infinite radius. 

\section{The main result} 
\setcounter{equation}{0}
In this section we give the precise  formulation of
our result and show how it is obtained from the implicit
function theorem, postponing the rather technical verification
of the assumptions of the latter to the last two sections.
We hope that most of our notation and terminology is self-explaining,
but the following needs to be introduced: The closed ball in $\R^3$ with 
center $0$ and radius $R>0$ is denoted by 
\[
B_R := \bigl\{ x \in \R^3 \mid \n{x} \leq R \bigr\},
\]
and
\[
\dot B_R := B_R \setminus \{0\}; 
\]
$\n{x}$ denotes the Euclidean norm of $x \in \R^3$. Also, let
\[
S_1 := \bigl\{ x \in \R^3 \mid \n{x} = 1 \bigr\},
\]
denote the unit sphere in $\R^3$, and denote the line segment joining 
two points $x, x' \in \R^3$ by
\[
\overline{x,\, x'} := \bigl\{ \lambda x + (1 - \lambda )x' \mid
\lambda \in [0,1] \bigr\} .
\]
The set of transformations which are to leave our solutions invariant is
\beas
S := \Bigl\{ A \in O(3) 
&\mid&
A \ \mbox{is a rotation around the $x_3$-axis}\\
&&
\mbox{or the mapping}\  \R^3 \ni (x_1,x_2,x_3) \mapsto (x_1,x_2,-x_3)\Bigr\}\, ;
\eeas
note that in addition to axial symmetry we require reflection 
symmetry with respect to the plane $\{ x_3 =0\}$.
Let
\[
C_S (B_R) := \Bigl\{ f \in C(B_R) \mid f(Ax) = f(x),\ A \in S,\ x \in B_R 
\Bigr\} .
\]
Clearly, 
\[
\nabla f (0) = 0,\ f \in C^1 (B_R) \cap C_S (B_R) ,
\]
and this is the reason for introducing the extra reflection symmetry.

For the phase space distribution function $f$ of our stationary solution 
we make the ansatz
\[
f (x,v) := \phi (E) \psi (\g P),\ x,v \in \R^3
\]
where $\g \in \R$ and $U$ is assumed to be  axially symmetric;
the quantities $E$ and $P$ were defined in (\ref{edef}) and (\ref{pdef}). 
Throughout
our investigation, $\phi : \R \to [0,\infty[$ and $\psi : \R \to [0,\infty[$
will satisfy the following assumptions:
\begin{itemize}
\item[($\phi 1$)]
$\phi\in L^p_{loc}(\R)$ for some $p>2$, and
there exists a constant $E_0 \in \R$ such that $\phi (E) = 0$ for $E \geq E_0$ a.~e.,
and
$\phi (E) >0$ for $E < E_0$ a.~e.
\item[($\phi 2)$]
The ansatz $f_0 (x,v) = \phi (E)$ leads to a nontrivial,  
static solution $(f_0,\r_0,U_0)$ of the Vlasov-Poisson system,
which is spherically symmetric, i.~e., $\r_0$ and $U_0$ depend only on
$\n{x}$, and such that
$ \r_0 \in C^1_c (\R^3)$ with $\supp \r_0 = B_1$ and $U_0 \in C^2 (\R^3)$
with $\lim_{\n{x} \to \infty} U_0 (x) = 0$.
\item[$(\psi )$]
$\psi \in C^2(\R)$ with $\psi' (0) = 0$ and $\psi (P) = 1 \Leftrightarrow P = 0$.
\end{itemize}
     
\noindent
{\bf Remark:}
For $E_0 \in \R$ and $-\frac{1}{2} < \mu < \frac{7}{2}$ the function
\[
\phi (E) := \left\{ \begin{array}{cll}
(E_0 - E)^\mu &,& E < E_0,\\
0 &,& E \geq E_0
\end{array} \right.
\]
obviously satisfies $(\phi 1)$
and leads to a spherically symmetric steady state $(f_0,\r_0,U_0)$
with finite radius and finite mass, cf.~\cite[Thm.\ 5.4]{BFH}. The solution 
has the required regularity, without loss of generality we can assume that
$\supp \r_0 = B_1$, and since $\lim_{\n{x} \to \infty} U_0 (x)$
exists we can take this limit to be zero by redefining $E_0$ accordingly.
Thus a large class of the so-called polytropic steady states satisfies
the assumptions $(\phi 1)$ and $(\phi 2)$.

\smallskip

\noindent {\bf Theorem:}
{\em There exists a constant $\g_0 >0$ such that for every 
$\g \in ]-\g_0,\g_0[$ there exists a nontrivial stationary solution 
$(f^\g,\r^\g,U^\g)$
of the Vlasov-Poisson system with the following properties:
\begin{itemize}
\item[{\rm (i)}]
$f^\g (x,v) = \phi (E) \psi (\g P),\ x,v \in \R^3$.
\item[{\rm (ii)}]
$(f^0,\r^0,U^0) = (f_0,\r_0,U_0)$, $(f^\g,\r^\g,U^\g)$ is axially symmetric
for $\n{\g} < \g_0$, more precisely, for all
$ x,v \in \R^3$ and $A \in S$ we have
\[
f^\g (Ax,Av) = f^\g (x,v),\ \r^\g (Ax) = \r^\g (x),\
U^\g (A x) = U^\g (x),
\]
and $(f^\g,\r^\g,U^\g)$ is not spherically symmetric for $\g \neq 0$, i.~e.,
the above identities fail if $S$ is replaced by $SO (3)$.
\item[{\rm (iii)}]
$\r^\g \in C^1_c (\R^3)$ and $U^\g \in C^2_b (\R^3)$ for 
$\n{\g} < \g_0$. 
\item[{\rm (iv)}]
The mappings $]-\g_0,\g_0[ \ni \g \mapsto \r^\g$ and
$]-\g_0,\g_0[ \ni \g \mapsto U^\g$ are continuous with respect to the 
norms $\nn{\cdot}_{1,\infty}$ or $\nn{\cdot}_{2,\infty}$ respectively.
\end{itemize} }

\smallskip

\noindent
The question whether the axially symmetric steady states obtained above
are static or not is addressed at the end of this section. 
In order to prove the above theorem we first deduce the semilinear 
elliptic problem (\ref{nopo}) introduced in the introduction:
\begin{lemma} \label{h}
Let $f$ be given by $f(x,v) := \phi (E) \psi (\g P)$ for some potential
$U:\R^3 \to \R$. Then 
\[
\rho (x) = h\bigl(\g,r(x),U(x)\bigr),\ x \in \R^3,
\]
where
\[
r(x) := \sqrt{x_1^2 + x_2^2},\ x \in \R^3,
\]
and  
\[
\renewcommand{\arraystretch}{2}
h(\g,r,u)  :=  
\left\{
\begin{array}{lll}
\displaystyle
2 \pi \int_u^{E_0} \phi (E) \int_{-\sqrt{2(E - u)}}^{\sqrt{2(E - u)}} 
\psi (\g r s)\, ds\, dE&,&\ u < E_0,\ \g \in \R,\ r \geq 0,\\
\displaystyle
0&,&\ u \geq E_0,\ \g \in \R,\ r \geq 0.
\end{array} \right.
\renewcommand{\arraystretch}{1}
\]
Moreover, $h \in C^1(\R \times [0,\infty[ \times \R)$, 
$\d{r} h \in C^1(\R \times [0,\infty[ \times \R)$, and for every 
bounded subset $B \subset \R \times [0,\infty[ \times \R$ there 
exist constants $C>0$ and $\nu \in ]0,1[$ with
\beas
\n{\d{r} h(\g,r,u)} 
&\leq&
C r,\\
\n{h(\g,r,u) - h(\g',r,u')}
&\leq& 
C \bigl( \n{\g - \g'}\, r + \n{u - u'} \bigr),\\
\n{\d{u} h(\g,r,u) - \d{u} h(\g',r,u')}
&\leq&
 C \bigl(\n{\g - \g'} + \n{u - u'}^\nu \bigr)
\eeas
for all $(\g,r,u),\ (\g',r,u') \in B$.
\end{lemma}
\prf
The formula for $h$ follows by introducing cylindrical coordinates
with respect to $(-x_2,x_1,0)/r(x)$ in velocity space; if $r(x) = 0$ 
then $\psi (\g P) =1$ and one can use spherical coordinates.
The function $h$ is easily seen to be continuously differentiable with
\beas
\d{\g} h(\g,r,u) 
&=& 
2 \pi r  \int_u^{E_0} \phi (E) \int_{-\sqrt{2(E - u)}}^{\sqrt{2(E - u)}} 
s \psi' (\g r s)\, ds\, dE,\\
\d{r} h(\g,r,u) 
&=& 
2 \pi \g  \int_u^{E_0} \phi (E) \int_{-\sqrt{2(E - u)}}^{\sqrt{2(E - u)}} 
s \psi' (\g r s)\, ds\, dE,\\
\d{u} h(\g,r,u) 
&=& 
- 2 \pi \int_u^{E_0}  
\Bigl(\psi \bigl(\g r \sqrt{2 (E - u)}\bigr) + 
\psi \bigl(- \g r \sqrt{2(E - u)}\bigr) \Bigr)
\frac{\phi (E)\, dE}{\sqrt{2(E - u)}}
\eeas
for $u < E_0$. The assumptions on $\psi$ imply that
$\n{\psi'(P)} \leq C \n{P}$ on bounded sets containing 0, 
which yields the estimate for $\d{r} h$. The second estimate is 
straightforward.  
Since  $\d{u} h$
is continuously differentiable with respect to $\g$, it is 
locally Lipschitz with respect to $\g$. 
As to the asserted H\"older continuity of $\d{u} h$ with respect to
$u$, take $(\g,r,u),\ (\g,r,u') \in B$, $B \subset \R \times [0,\infty[\times \R$
bounded, assume $u \leq u'$, and let $\frac{1}{p}+\frac{1}{q}=1$.
Then
\beas
\left|\d{u} h(\g,r,u) - \d{u}h(\g,r,u')\right|
&\leq&
C\, \int_u^{u'} \frac{\phi(E)}{\sqrt{E-u}}dE\\
&&
+ C\, \int_{u'}^{E_0} \n{\sqrt{E-u}-\sqrt{E-u'}}\, \frac{\phi(E)}{\sqrt{E-u}}dE\\
&&
+ C\, \left| \frac{1}{\sqrt{E-u}}-\frac{1}{\sqrt{E-u'}}\right|\, \phi(E)\, dE\\
&\leq&
C\, \n{u-u'}^{\frac{1}{q}-\frac{1}{2}}
+ C\, \n{u-u'}^{\frac{1}{2}} + 
C\, \n{u-u'}^{\frac{1}{q}-\frac{1}{2}},
\eeas
where we have used H\"older's inequality for the first and the last term.
Since $\frac{1}{q}-\frac{1}{2} > 0$ this completes the proof.
\prfe 
We note that the above estimates would simplify if we asssumed that $\phi$
is H\"older continuous, but this would exclude the polytropes with
$-\frac{1}{2} < \mu \leq 0$.
Next we collect some further properties
of the spherically symmetric steady state  $(f_0,\r_0,U_0)$:
\begin{lemma} \label{ss}
The spherically symmetric steady state $(f_0,\r_0,U_0)$
has the following additional properties:
\begin{itemize}
\item[{\rm (a)}]
The function $h(0, \cdot,\cdot)$ does not depend on the variable $r(x)$;
we will write it as $h_0 = h_0 (u)$ for simplicity.
\item[{\rm (b)}]
The potential $U_0$ is given by
\[
U_0 (x) = - \int \frac{\r_0 (y)}{\n{x - y}} dy
= - \frac{4 \pi}{\n{x}} \int_0^{\n{x}} s^2 \r_0 (s) \, ds
- 4 \pi \int_{\n{x}}^\infty s \r_0 (s) \, ds,\ x \in \R^3,
\]
and
\[
U_0' (\n{x}) = \frac{4 \pi}{\n{x}^2} 
\int_0^{\n{x}} s^2 \r_0 (s) \, ds ,\ x \in \R^3.
\]
\item[{\rm (c)}]
$\r_0$ is decreasing with $\r_0 (0) >0$, $U_0''(0) >0$,
for every $R>0$ there exists $C>0$ such that $U_0'(r) \geq C r,\ r\in [0,R]$,
and $U_0 (1) = E_0$.
\item[{\rm (d)}] 
$\r_0'$ is H\"older continuous, and $U_0' \in C^2(\rdd)$  
where $\rdd := \R^3 \setminus \{0\}$.
\end{itemize}
We identify $\r_0$ and $U_0$ as functions of $\n{x}$ with 
$\r_0$ and $U_0$ as functions 
of $x$; the derivative with respect to $\n{x}$ is denoted by $'$.
\end{lemma}
\prf
The assertion in (a) is obvious from Lemma~\ref{h}.
Since we require that $\lim_{\n{x} \to \infty} U_0(x) = 0$, 
the assertion in (b) holds by uniqueness.
Since $h_0$ is decreasing and $U_0$ is increasing
we find that $\r_0$ is also decreasing,
and since the steady state $(f_0,\r_0,U_0)$ is assumed to be nontrivial
we must have $\r_0 (0) >0$. Thus actually $U_0' (r) >0,\ r>0$,
and since $U_0''(0) = \frac{4 \pi}{3} \r_0 (0) >0$ this implies
the estimate on $U_0'$ from below. The assertion that $U_0(1) = E_0$
follows from the form of $h_0$ and the fact that by assumption 
$\supp \r_0 = B_1$. 
The regularity of $U_0'$ follows from the formula above 
and the fact that $\r_0 \in C^1_c (\R^3)$. 
Finally,  the H\"older continuity of  
$\r_0' (r) = \d{u} h_0 (U_0 (r))\, U_0'(r)$ 
follows from Lemma~\ref{h}. \prfe 
We want to find solutions of the equation
\be \label{nop}
\lap U = 4 \pi h(\g,r(x),U),
\ee
and the central idea is to reformulate this as a problem of finding
zeros of an operator $T$ which acts not on the space of potentials directly
but on deformations 
of the given spherically symmetric potential $U_0$.
We now define the Banach spaces which will serve as domain and range
of $T$:
\beas
&&
X := \Bigl\{ f \in C_S (B_3)  \mid 
f(0) = 0,\ f \in C^1 (\dot B_3),\ \exists C>0: \n{\nabla f (x)} \leq C,\ 
x \in \dot B_3,\\
 && \hspace{80pt}
\forall x \in S_1:\; \lim_{t \searrow 0} \nabla f(t x) =: \nabla f (0 x)
\ \mbox{exists, uniformly in}\ x \in S_1 \Bigr\},
\eeas
which we equip with the norm
\[
\nn{f}_X := \sup_{x \in \dot B_3} \n{\nabla f(x)},\ f \in X ,
\]
and
\beas
&&
Y := \Bigl\{ f \in C_S (B_3) 
\mid 
f(0) = 0,\ f \in C^1 (B_3),\ 
\exists C>0 : \n{\nabla f (x)} \leq C \n{x},\ x \in B_3,\\
&& \hspace{75pt} 
\forall x \in S_1:\ \lim_{t \searrow 0} \frac{\nabla f(t x)}{t} =: 
\frac{\nabla f (0 x)}{0}\ \mbox{exists, uniformly in}\ x \in S_1 \Bigr\},
\eeas 
which we equip with the norm
\[
\nn{f}_Y := \sup_{x \in \dot B_3} \frac{\n{\nabla f(x)}}{\n{x}},\ f \in Y .
\]  
For $f \in X$ the function $\nabla f(0 \,\cdot\,)$, being defined as the 
uniform limit of functions in $C(S_1)$, is itself in $C(S_1)$. 
Furthermore, since $f(0) = 0$, 
\[
\n{f(x)} \leq \nn{f}_X \n{x},\ x \in B_3,\ f \in X,
\]
and the norm $\nn{\cdot}_X$ is equivalent to the norm
\[
||| f |||_X := \sup_{x \in \dot B_3} 
\left( \frac{\n{f(x)}}{\n{x}} + \n{\nabla f(x)} \right) + 
\sup_{x\in S_1} \n{\nabla f(0 x)} .
\]
It easily follows that $(X, \nn{\cdot}_X)$ is a Banach space.
For $f \in Y$ note first that
\[
\frac{f (0 x)}{0^2} := \lim_{t \searrow 0} \frac{f(t x)}{t^2} = 
\frac{1}{2} \frac{\nabla f (0 x)}{0} \cdot x
\]
uniformly in $x \in S_1$. We have
\[
\frac{\nabla f(0\,\cdot\,)}{0},\ \frac{f(0\,\cdot\,)}{0^2} \in C (S_1),\ \ 
\n{f(x)} \leq \nn{f}_Y \n{x}^2,\ x \in B_3,
\]
and the norm $\nn{\cdot}_Y$ is equivalent to the norm
\[
||| f |||_Y := \sup_{x \in \dot B_3} 
\left( \frac{\n{f(x)}}{\n{x}^2} + \frac{\n{\nabla f(x)}}{\n{x}} \right) + 
\sup_{x\in S_1} \left(\left|\frac{f(0x)}{0^2}\right| + 
\left|\frac{\nabla f(0 x)}{0}\right| \right) ,
\]
from which it follows that $(Y, \nn{\cdot}_Y)$ is a Banach space.

Using the elements in the Banach space $X$ we can deform spherically
symmetric sets, e.~g., the level sets of the given, spherically
symmetric static solution, into axially symmetric sets in the following way:
\begin{lemma} \label{defo}
For $\z \in X$ define
\[
g_\z : B_3 \to \R^3,\
g_\z (x) := x + \z (x) \frac{x}{\n{x}},\ x \in \dot B_3,\ g_\z (0) := 0 .
\]
Then there exists $r>0$ such that for all
\[
\z \in \Omega := \bigl\{ \z \in X \mid \nn{\z}_X < r \bigr\} 
\]
the following holds:
\begin{itemize}
\item[{\rm (a)}]
$g_\z : B_3 \to B_{3,\z} := g_\z (B_3)$ is a homeomorphism,
$g_\z : \dot B_3 \to \dot B_{3,\z}$ is a $C^1$-diff\-e\-o\-morphism
whose Jacobian satisfies the estimate
\[
\n{D g_\z (x) - id} < \frac{1}{2},\ x \in \dot B_3,
\]
and for every $x \in S_1$ the restriction 
\[
g_\z : \overline{0,\, 3x} \ni y \mapsto g_\z (y) \in 
\overline{0,\,\n{g_\z (3 x)}x}
\]
is one-to-one, onto, and preserves
the natural ordering of points on the line segment $\overline{0,\,3 x}$. 
\item[{\rm (b)}]
$\displaystyle 
\frac{1}{2} \n{x} \leq \n{g_\z (x)} \leq \frac{3}{2} \n{x},\ x \in B_3$, and
$\displaystyle
g_\z (B_1) \subset \open{B}_2,\ B_2 \subset g_\z(B_3) \subset B_4$.
\item[{\rm (c)}]
$g_\z (A x) = A g_\z (x),\ x \in B_3$, and 
$g_\z^{-1} (A x) = A g_\z^{-1} (x),\ x \in B_{3,\z},\ A \in S$.
\item[{\rm (d)}]
$\displaystyle 
\n{Dg_\z^{-1} (x) - id} < \frac{1}{2},\ x \in \dot B_{3,\z}$,
and there exists a constant $C>0$ such that for all
$\z, \z'\in \O$,
\[
\frac{1}{\n{x}}\n{g_\z (x) - g_{\z'} (x)} +
\n{D g_\z (x) - D g_{\z'} (x)} \leq C \nn{\z - \z'}_X ,\ x \in \dot B_3,
\]
and
\[
\n{g_\z^{-1} (x) - g_{\z'}^{-1} (x)}
\leq C \nn{\z - \z'}_X \n{x} ,\ x \in B_2 .
\]
\end{itemize}
\end{lemma}
\prf
On $\dot B_3$ we have for $i,j =1,2,3$,
\be \label{dg}
\d{x_i} g_{\z,j} (x) = \delta_{ij} + \d{x_i} \z (x) \frac{x_j}{\n{x}}
+ \frac{\z (x)}{\n{x}} \left( \delta_{ij} - \frac{x_i x_j}{\n{x}^2} \right)
\ee
which implies that 
\[
\n{D g_\z (x) - id} \leq 3 \nn{\z}_X < \frac{1}{2},\ x \in \dot B_3,
\] 
provided $\nn{\z}_X < 1/6$. Using the inverse function theorem
we obtain the first two assertions in (a). Since $g_\z (y) \in ]0,\infty[\, y$
for every $y \in B_3$ the remaining assertion in (a) follows. The assertions in (b) 
are obvious, provided $r>0$ is chosen
sufficiently small, and so are the assertions in (c). 
As to (d), the first assertion follows by choosing $r$ still smaller,  
since $D g_\z ^{-1} (x) = (D g_\z)^{-1} (g_\z^{-1} (x))$ 
The estimate for $g_\z - g_{\z'}$ 
is immediate from the definition of $g_\z$.  The estimate for 
$D g_\z - D g_{\z'}$ follows from (\ref{dg}). Finally,
$x \in \dot B_2$ implies that $x \in g_\z (B_3) \cap g_{\z'} (B_3)$,
and there exists $y \in \dot B_3$ such that $x = g_{\z'} (y)$. Thus
\beas
\n{g_\z^{-1} (x) - g_{\z'}^{-1} (x)}
&=&
\n{g_\z^{-1} (g_{\z'} (y)) -y} =
\n{ g_\z^{-1} (g_{\z'} (y)) - g_\z^{-1} (g_\z (y))}\\
&\leq&
2 \n{g_\z (y) - g_{\z'} (y)} \leq 2 \nn{\z - \z'}_X \n{y}
\leq
4 \nn{\z - \z'}_X \n{x}
\eeas
by the mean value theorem, the estimate for $D g_\z^{-1}$ which we already 
established and the fact that 
$\overline{g_\z (y),\,g_{\z'} (y)} \subset g_\z (\dot B_3)$. \prfe
We want to find solutions of the reduced problem (\ref{nop})
of the form
\[
U(x) = U_\z (x) := U_0 (g_\z^{-1} (x)),\ x \in B_{3,\z},
\]
for some $\z \in \O$. 
Of course $U$ will have to be defined on all of $\R^3$, but this will be
easy once we have it on $B_{3,\z}$.
Using the fundamental solution of the Poisson equation
we integrate (\ref{nop}) and
transform our problem  to that of solving the equation
\[
U_0 (x) + \int_{B_{3,\z}} 
\frac{h \bigl(\g,r(y), U_0 (g_\z^{-1}(y))\bigr)}{\n{g_\z (x) - y}} dy =0,\
x \in B_3;
\]
observe that $g_\z$ is invertible.
It turns out that we can avoid the dependence of the domain of
integration on $\z$, and also that the operator above is not quite 
the right thing yet. We are now in the position to
give the proof of the theorem:
 
\smallskip

\noindent
{\bf Proof of the Theorem:}
For $\z \in \Omega$ and $\g \in \R$ we define 
\bea
T(\g,\z) (x) 
&:=&
U_0 (x) + \int_{B_2} 
\frac{h \bigl(\g,r(y), U_0 (g_\z^{-1}(y))\bigr)}{\n{g_\z (x) - y}} dy 
\nonumber \\
&&
\mbox{} - U_0 (0) - \int_{B_2} 
\frac{h \bigl(\g,r(y), U_0 (g_\z^{-1}(y))\bigr)}{\n{y}} dy,\ x \in B_3.
\label{tdef}
\eea
Assume we already know
that this defines a continuous operator
$T : ]-1,1[ \times \Omega \to Y$, that $T$ is continuously Fr\'{e}chet 
differentiable with respect to $\z$, and that
\[
\d{\z} T(0,0) : X \to Y
\]
is an isomorphism; that all this is indeed the case is shown in 
Sections 3 and 4, cf.\ Proposition~\ref{frechet} and Proposition~\ref{isom}.
Note that by definition of $Y$ we must have 
$T(\g,\z)(0) = 0$
which is the reason for subtracting the constant term in the definition of $T$. 
By assumption $(\phi 2)$ we know that $T(0,0) = 0$;
note that $g_0 = id$ and that 
$\supp \r_0 = \supp h_0 \circ U_0 = B_1 \subset B_2$. 
By the implicit function theorem there exists a constant 
$\g_0 \in ]0,1[$ and a continuous mapping
\[ 
]-\g_0,\g_0[ \ni \g \mapsto \z^\g \in \Omega
\]
such that
\[
T(\g,\z^\g) = 0,\ \g \in ]-\g_0, \g_0[
\]
and $\z^0 = 0$, cf.\ \cite[Theorem 15.1]{D}. 
Let $\z = \z^\g$ for some $\g \in ]-\g_0,\g_0[$ and define
\be \label{rzdef}
\r_\z (x) := h\bigl(\g, r(x), U_0 (g_\z^{-1}(x))\bigr),\ x \in B_2 .
\ee
Then $\r_\z \in C_S (B_2)$; at the moment the differentiability at $x=0$
is not yet obvious. By Lemma~\ref{h}
$\r_\z (x) > 0$ if and only if $U_0 (g_\z^{-1} (x)) < E_0$
which by Lemma~\ref{ss}
is equivalent to $\n{g_\z^{-1} (x)} <1$. Therefore, by Lemma~\ref{defo}
\[
\supp \r_\z = g_\z(B_1) \subset \open{B}_2,
\]
and we can extend $\r_\z$ by 0 to all of $\R^3$, obtaining an element of
$C_c (\R^3)$ with $\supp \r_\z \subset \open{B}_2$ which for the moment
need not satisfy (\ref{rzdef}) everywhere. The equation
$T(\g, \z) = 0$ can now be written as
\[
U_0 (x) = - \int_{B_2} \frac{\r_\z (y)}{\n{g_\z (x) - y}} dy\, + C,\ x \in B_3,
\]
or
\[
U_0 \bigl(g_\z ^{-1} (x)\bigr) = 
-  \int_{B_2} \frac{\r_\z (y)}{\n{x - y}} dy\, + C,\ x \in B_{3,\z},
\]
where
\[
C:= U_0(0) +  \int_{B_2} \frac{\r_\z (y)}{\n{y}} dy .
\]
Now define
\[
U_\z (x) := - \int_{\R^3} \frac{\r_\z (y)}{\n{x - y}} dy\, + C,\ x \in \R^3 .
\]
Clearly, $U_\z \in C^1 (\R^3)$ with 
\be \label{uzeta}
U_\z (x) = U_0 \bigl(g_\z ^{-1}(x)\bigr),\  x \in B_2 \subset B_{3,\z}.
\ee
This immediately implies that $\r_\z \in C^1_c (\R^3)$ and  
$U_\z \in C^2_b (\R^3)$ with $\lap U_\z = 4 \pi \r_\z$ on $\R^3$.
On the other hand,
\be \label{nopoz}
\lap U_\z (x) = 4 \pi h \bigl(\g, r(x), U_\z (x)\bigr),\ x \in B_2 \subset B_{3,\z}.
\ee
If fact the latter equation holds on all of $\R^3$. To see this we have to 
show that $\r_\z (x) = h \bigl(\g, r(x), U_\z (x)\bigr),\ x \in \R^3$, that is,
$U_\z (x) > E_0$ for $x \in \R^3 \setminus g_\z (B_1)$.
We know that 
\[
\lap U_\z (x) = 0,\ x \in \R^3 \setminus g_\z (B_1), 
\]
and
\[
\lim_{\n{x} \to \infty} U_\z (x) = C,\quad
U_\z (x) = E_0,\ x \in \partial g_\z (B_1),\quad 
U_\z (x) > E_0,\ x \in B_2\setminus g_\z (B_1), 
\]
where we used the identity (\ref{uzeta})
and the fact that $U_0$ is strictly increasing as a function of $\n{x}$
with $U_0 (1) = E_0$. The assumption $C \leq E_0$ would  
contradict the maximum principle. Thus, $C> E_0$, and again by the 
maximum principle,
$U_\z > E_0$ on $\R^3 \setminus g_\z (B_1)$. Therefore, (\ref{nopoz})
does hold on all of $\R^3$.

If we define $\r^\g := \r_{\z_\g}$, $U^\g := U_{\z_\g}$, and
\[
f^\g (x,v) := \phi \left(\frac{1}{2} v^2 + U^\g (x) \right)
               \psi \bigl( \g \,(x_1 v_2 - x_2 v_1) \bigr),\ x,v \in \R^3,
\]
then the assertions (i)--(iii) of the theorem are established,
except for the assertion that the solution is not spherically symmetric
for $\g \neq 0$. To see the latter, choose $x \in \R^3$ with
$\r^\g (x) >0$, $x_1 \neq 0,\ x_2 = x_3 = 0$. There must then
exist $\eta \neq 0$ such that $\frac{1}{2} \eta^2 + U^\g (x) < E_0$.
Let $v =(0,0,\eta)$ and $v' = (0,\eta,0)$. Then there exists
a rotation $A$ around the $x_1$-axis such that $A v = v'$, 
and clearly, $A x = x$. But since $E(x,v) = E(x,v')$ and 
$P(x,v)=0 \neq x_1 \eta = P(x,v')$ we have
\beas
f^\g (x,v)  
&=&
\phi \bigl(E(x,v)\bigr) \psi \bigl(\g P(x,v)\bigr) =
\phi \bigl(E(x,v)\bigr) = \phi \bigl(E(x,v')\bigr)\\
&\neq&
\phi\bigl(E(x,v')\bigr) \psi \bigl(\g P(x,v')\bigr)
=
f^\g (x,v'),
\eeas
provided $\g \neq 0$.
The continuity properties asserted in (iv) follow from the fact that
$\z_\g$ depends continuously on $\g$ with respect to the norm $\nn{\cdot}_X$.
First the estimate
\[
\n{U^\g (x) - U^{\g'} (x)}
\leq
\nn{U_0'}_\infty \bigl| g_{\z_\g}^{-1} (x) - g_{\z_{\g'}}^{-1} (x)\bigr|\\
\leq
C\, \nn{\z_\g - \z_{\g'}}_X,\ x \in B_2,
\]
and the relation $\r^\g (x) = h(\g,r(x),U^\g (x))$ imply that
$\r^\g$ depends continuously on $\g$ with respect to $\nn{\cdot}_\infty$.
Since
\[
U^\g (x) = - \int_{B_2} \frac{\r^\g (y)}{\n{x - y}} dy + U_0(0)
+ \int_{B_2} \frac{\r^\g (y)}{\n{y}} dy,\ x \in \R^3,
\]
this implies that $U^\g$ depends continuously on $\g$ with respect to
$\nn{\cdot}_{1,\infty}$.
Differentiating the above formula for $\r^\g$ we obtain the asserted
continuity of $\r^\g$ with respect to $\nn{\cdot}_{1,\infty}$
and thus also of $U^\g$ with respect to
$\nn{\cdot}_{2,\infty}$, and the proof of the theorem is complete.
\prfe 
\noindent
{\bf Remark:}\nopagebreak\begin{itemize}
\item[(a)]
For fixed $\psi$ the family $(f^\g,\r^\g,U^\g)$ is 
in a neighbourhood of $\g =0$ unique. However,
different functions $\psi$ give different families of stationary solutions
in which $(f_0,\r_0,U_0)$ is embedded.
\item[(b)]
The mass current density is given by 
\beas
j^\g (x) 
&:=& 
\int v f^\g (x,v) \, dv \\
&=&
2 \pi \int_{U^\g (x)}^{E_0}
\phi (E) \int_{-\sqrt{2 (E - U^\g (x))}}^{\sqrt{2 (E - U^\g (x))}}
s \psi (\g r(x) s)\, ds\, dE\; e_t (x),\ x \in \R^3,
\eeas
where
\[
 e_t (x) := \frac{1}{r(x)} (-x_2,x_1,0)
\]
denotes the unit vector field tangent to the orbits of points under 
counterclockwise rotations
around the $x_3$-axis; note that the integral above 
vanishes on the $x_3$-axis, where $e_t$ is not defined.
Now the average velocity $j^\g/\r^\g$ of the steady state vanishes
identically
if $\psi$ is an even function, in which case we obtain a family
of static, axially symmetric solutions. In general, for example
if $\psi (-P) < 1 < \psi (P),\ P >0$, the average velocity does not vanish,
and  we obtain a stationary stellar system which 
rotates around the $x_3$-axis.
It is also easy to see that the average velocity vanishes
at the boundary of the support of $\r^\g$.  
As already mentioned in the introduction,
static solutions of the Euler-Poisson system
are necessarily spherically symmetric.
As we now see, a corresponding result does not hold for the Vlasov-Poisson 
system. 
\item[(c)]
It would require very little additional efford to show
that $T$ is continuously differentiable also with respect to $\g$. 
This would imply 
that the family $(f^\g,\r^\g,U^\g)$ depends on $\g$ in a differentiable way
with respect to the appropriate norms, cf.\ \cite[Cor.~15.1]{D}.
\item[(d)]
If in our ansatz $f$ depends only on the particle energy $E$ then the 
right hand side of (\ref{nopo}) does not explicitly depend on $x$.
One can then apply a result by {\sc Gidas, Ni}, and {\sc Nirenberg},
cf.\ \cite[Theorem 4]{GNN}, to conclude that under mild regularity
assumptions a corresponding steady state with finite radius and finite
mass must be spherically symmetric with respect to some point in $\R^3$.
Therefore, it is necessary to include further invariants in the ansatz
in order to obtain stationary models which are not spherically symmetric.
\end{itemize}
\section{The Fr\'{e}chet-differentiability of $T$}
\setcounter{equation}{0}
The aim of this section is to prove the following result:
\begin{prop} \label{frechet}
The mapping $T : ]-1,1[ \times \O \to Y$ defined in (\ref{tdef})
is continuous and continuously Fr\'{e}chet-differentiable with respect
to $\z$ with Fr\'{e}chet derivative
\beas
&&
\bigl[\d{\z} T(\g,\z) \x\bigr] (x) \\
&&
\hspace{8pt}=
- \int_{B_2} \left( \frac{1}{\n{g_\z (x) - y}} - \frac{1}{\n{y}} \right) 
\d{u} h\bigl(\g,r(y),U_\z(y)\bigr) \nabla U_\z (y)
\cdot \frac{g_\z^{-1} (y)}{\n{g_\z^{-1} (y)}} 
\x \bigl(g_\z^{-1} (y)\bigr)\, dy\\
&&
\hspace{8pt} \quad - \int_{B_2} \frac{g_\z (x) - y}{\n{g_\z (x) - y}^3}
h\bigl(\g, r(y), U_\z (y)\bigr) \, dy \cdot \frac{x}{\n{x}} \x (x),\ 
x \in B_3,
\eeas
where $\g \in ]-1,1[$, $\z \in \O$, $\x \in X$, and 
$U_\z (y) := U_0 (g_\z^{-1} (y)),\ y \in B_2$.
\end{prop}
In order to prove this result we need more information on 
the elements of the space $X$
and the deformation mappings constructed in Lemma~\ref{defo}. 
\begin{lemma} \label{zeta}
Let $\z \in \O$. Then the following holds: 
\begin{itemize}
\item[{\rm (a)}]
$ \n{\z (x) - \z (x')} \leq \nn{\z}_X \n{x - x'},\ x, x' \in B_3$.
\item[{\rm (b)}]
For $x \in S_1$ the mapping $[0,3] \ni t \mapsto \z (t x)$ is 
continuously differentiable,
and $\lim_{t \searrow 0} \frac{\z (tx)}{t}$ $= \nabla \z (0 x)\cdot x$, 
uniformly
in $x \in S_1$.
\item[{\rm (c)}]
The mapping $[0,3] \times S_1 \ni (t,x) \mapsto \nabla \z (t x)$
is uniformly continuous.
\item[{\rm (d)}]
For $x \in S_1$ the limits 
$\lim_{t \searrow 0} \frac{g_\z (t x)}{t} =:\frac{g_\z (0 x)}{0}$
and
$\lim_{t \searrow 0} D g_\z (t x) =: D g_\z (0 x)$ exist,
uniformly in $x \in S_1$.
\end{itemize}
\end{lemma}
\prf
The assertion in (a) follows easily by distinguishing the cases
$0 \in \overline{x,\, x'}$ and $0 \not \in \overline{x,\, x'}$.
As to (b),
\[
\frac{d}{dt} \z (t x) = \nabla \z (t x) \cdot x \to  \nabla \z (0 x) 
\cdot x,\
t \searrow 0,
\]
by definition of the space $X$, and the rest follows. The assertion in (c)
follows from the fact that $\nabla \z \in C(\dot B_3)$ and 
$\nabla \z (t x) \to \nabla \z (0 x)$ uniformly in $x \in S_1$
as $t \searrow 0$. The assertions in (d) are easy consequences of
the definitions of $g_\z$ and the space $X$, together with (\ref{dg}).
\prfe 
Next we establish some estimates for the spatial density induced by a
deformation of the potential $U_0$:
\begin{lemma} \label{rho}
For $\g \in ]-1,1[$ and $\z \in \O$ let
\[
\r_{\g,\z} (x) := h\bigl(\g,r(x), U_0(g_\z ^{-1} (x))\bigr),\ x \in B_2.
\]
Then the following holds:
\begin{itemize}
\item[{\rm (a)}]
$\r_{\g,\z} \in C_S (B_2) \cap C^1 (B_2)$ with
$\supp \r_{\g,\z} \subset \open{B}_2$,
and there exists a constant $C>0$ such that for all $\g \in ]-1,1[$ 
and $\z \in \O$,
\[
\n{\nabla \r_{\g,\z} (x)} \leq C \n{x},\ x \in B_2 .
\]
\item[{\rm (b)}]
There exists a constant $C>0$ such that for all $\g,\g' \in ]-1,1[$ and 
$\z,\z' \in \O$,
\[
\n{\r_{\g,\z} (x) - \r_{\g',\z'} (x) } \leq C 
\Bigl(\n{\g - \g'} + \nn{\z - \z'}_X \Bigr) \n{x},\ x \in B_2 .
\]
\end{itemize}
\end{lemma}
\prf
Lemma~\ref{defo} and Lemma~\ref{h} 
imply that $\r = \r_{\g,\z} \in C_S (B_2) \cap C^1 (\dot B_2)$.
For $x \in \dot B_2$ we have
\beas
\nabla \r_{\g,\z} (x) 
&=&
\d{r} h\bigl(\g,r(x),U_0(g_\z^{-1}(x))\bigr) \nabla r(x) \\
&&
\mbox{}+   
\d{u} h\bigl(\g,r(x),U_0(g_\z^{-1}(x))\bigr)
\nabla U_0 (g_\z^{-1}(x)) \cdot D g_\z^{-1} (x),
\eeas
and Lemma~\ref{h}, the fact that $U_0 \in C^2(\R^3)$ with
$\nabla U_0 (0) = 0$, and Lemma~\ref{defo} 
imply the estimate
\[
\n{\nabla \r (x)} \leq
C \n{x} + C \n{\nabla U_0 (g_\z^{-1} (x))} \leq
C \n{x} + C \n{g_\z^{-1} (x)} \leq C \n{x},\ x \in \dot B_2;
\]
note that the range of $U_0$ is bounded. Since $x \not \in g_\z (B_1)$ 
implies $U_0 (g_\z^{-1}(x)) >E_0$
and thus $\r (x) = 0$, the assertion on the support of $\r$
follows by Lemma~\ref{defo}~(b).
The assertion in (b) is immediate from Lemma~\ref{h} 
and Lemma~\ref{defo}~(d). \prfe
We shall need the following assertions on Newtonian potentials:
\begin{lemma} \label{newton}
Let $\s \in C_S (B_2)$ be such that 
\[
c_\s := \sup_{x \in \dot B_2} \frac{\n{\s (x)}}{\n{x}} < \infty 
\]
and define
\[
V_\s (x) := - \int_{B_2} \frac{\s (y)}{\n{x - y}} dy,\ x \in \R^3 .
\]
Then $V_\s \in C^1 (\R^3)$, and there exists $C>0$ such that for all
$\s$ as above the following estimates hold:
\begin{itemize}
\item[{\rm (a)}]
$\displaystyle \qquad \n{\nabla V_\s (x)} \leq C c_\s \n{x},\ x \in \R^3$,
\item[{\rm (b)}]
$\displaystyle \qquad 
\bigl|\nabla V_\s (g_\z (x)) - \nabla V_\s (g_{\z'} (x))\bigr| 
\leq C c_\s \nn{\z - \z'}_X^{1/2} \n{x},\ x \in B_3,\ \z, \z' \in \O$.
\end{itemize}
\end{lemma}
\prf
For $\s \in C_S(B_2)$ we have $\nabla V_\s (0) = 0$ and thus
\[
\n{\nabla V_\s (x)}
\leq
\left| \int_{B_2} \left( \frac{x-y}{\n{x-y}^3} + \frac{y}{\n{y}^3}\right)
\s (y)\, dy \right| ,\ x \in \R^3.
\]
Let $x \neq 0$ and $r := 2 \n{x}$. We obtain the estimate
\beas
\n{\nabla V_\s (x)} 
&\leq&
c_\s \int_{B_2\setminus B_r} \left| \frac{x-y}{\n{x-y}^3} + 
\frac{y}{\n{y}^3}\right|
\n{y}\, dy  + 
c_\s \int_{B_2\cap B_r} \left( \frac{1}{\n{x-y}^2} + \frac{1}{\n{y}^2}\right)
\n{y}\, dy\\
&=:&
I_1 + I_2.
\eeas
For almost every $y \in B_2$ there exists $\tau \in [0,1]$ such that
\[
\left| \frac{x-y}{\n{x-y}^3} + \frac{y}{\n{y}^3}\right|
\leq \n{x} \frac{4}{\n{\tau x - y}^3}
\]
and since for $\n{y} \geq r$, 
\[
\n{\tau x - y} \geq \n{y} - \n{x} = \n{y} - \frac{r}{2} \geq \frac{\n{y}}{2},
\]
we can estimate the first term as
\[
I_1 \leq C c_\s \n{x} \int_{B_2} \frac{1}{\n{y}^2} dy = C c_\s \n{x} ;
\]
constants denoted by $C$ may change their value from line to line or 
even within one and the same line. For the second term we have
\[
I_2 \leq
2 c_\s \left( \int_{B_r} \frac{1}{\n{x - y}^2} dy
+  \int_{B_r} \frac{1}{\n{y}^2} dy \right)
\leq 
4 c_\s \int_{B_r} \frac{1}{\n{y}^2} dy = C c_\s r = C c_\s \n{x},
\]
and the proof of part (a) is complete. As to (b) we have
\[
\bigl|\nabla V_\s (g_\z (x)) - \nabla V_\s (g_{\z'}(x))\bigr|
\leq
c_\s \int_{B_2} 
\left| \frac{g_\z (x) - y}{\n{g_\z (x) - y}^3}
- \frac{g_{\z'} (x) - y}{\n{g_{\z'} (x) - y}^3} \right| \n{y}\, dy .
\]
Let $x \in \dot B_3$ and $\delta:= \nn{\z - \z'}_X <1$,
$r_1 := 2 \delta \n{x}$, and $r_2 := 4 \n{x} > r_1$;
recall that we chose the radius of the set $\O$ less than $1/3$.
We split the integral above into three parts, $I_1,\ I_2$, and $I_3$, 
according to the decomposition
\[
B_2 = \Bigl( B_2 \setminus B_{r_2} \Bigr) \cup
\Bigl( \bigl(B_2 \cap B_{r_2} \bigr) \setminus B_{r_1}(g_\z (x))\Bigr)
\cup \Bigl( B_2 \cap B_{r_1}(g_\z (x))\Bigr) .
\]
As to $I_1$ we find for almost every $y \in B_2$ a $\tau$ between
$\z (x)$ and $\z' (x)$ such that
\[
\left| \frac{g_\z (x) - y}{\n{g_\z (x) - y}^3}
- \frac{g_{\z'} (x) - y}{\n{g_{\z'} (x) - y}^3} \right|
\leq \frac{C}{\n{ x + \tau \frac{x}{\n{x}} - y}^3}
\n{\z (x) - \z' (x)};
\]
note that both $g_\z (x)$ and $g_{\z'} (x)$ lie on the line $\R x$.
Since 
\[
\n{\z (x) - \z' (x)} \leq \nn{\z - \z'}_X \n{x} = \delta \n{x},
\]
and for $\n{y} \geq r_2$,
\beas
\left|x + \tau \frac{x}{\n{x}} - y\right| 
&=&
\left| y - g_\z (x) + (\z (x) - \tau) \frac{x}{\n{x}}\right|
\geq
\n{y} - \n{g_\z (x)} - \n{\z (x) - \z' (x)}\\
&\geq&
\n{y} - \frac{3}{2} \n{x} - \delta \n{x} \geq \n{y} - \frac{5}{2} \n{x}
=
\n{y} - \frac{5}{8} r_2 \geq \frac{3}{8} \n{y},
\eeas
we find the estimate
\[
I_1
\leq
C c_\s \nn{\z - \z'}_X \n{x} \int_{B_2} \frac{dy}{\n{y}^2}
=
C c_\s \nn{\z - \z'}_X \n{x} .
\]
To estimate the second term $I_2$ we start like for $I_1$, but for 
$y \notin B_{r_1}(g_\z(x))$ obtain the estimate
\beas
\left| x + \tau \frac{x}{\n{x}} - y\right|
&=&
\left| y - g_\z (x) + (\z (x) - \tau) \frac{x}{\n{x}}\right| 
\geq
\n{y - g_\z (x)} - \n{\z (x) - \z'(x)}\\
&\geq&
\n{y - g_\z (x)} - \delta \n{x} \geq \frac{1}{2}\n{y - g_\z (x)} .
\eeas
On the other hand for $y \in B_{r_2}$ we have
\[
\n{y - g_\z (x)} \leq \n{y} + \frac{3}{2} \n{x} \leq r_2 + \frac{3}{8} r_2
\leq 2 r_2,
\]
and
\beas
I_2
&\leq&
C c_\s \delta \n{x} \int_{B_{2 r_2} (g_\z (x)) \setminus B_{r_1}(g_\z (x))}
\frac{1}{\n{ g_\z (x) - y}^3} dy  =
C c_\s \delta \n{x}\, 4 \pi \ln \frac{2 r_2}{r_1} \\ 
&=&
C c_\s \delta \n{x}\, \ln\frac{4}{\delta} 
\leq
C c_\s \delta^{1/2} \n{x} = C c_\s \nn{\z - \z'}_X^{1/2} \n{x} .
\eeas
As to the third term we have
\beas
I_3 
&\leq&   
2 c_\s \left( \int_{B_{r_1} (g_\z (x))} \frac{dy}{\n{g_\z (x) - y}^2}
+ \int_{B_{r_1} (g_\z (x))} \frac{dy}{\n{g_{\z'} (x) - y}^2} \right) \\
&\leq&
4 c_\s \int_{B_{r_1} (g_\z (x))} \frac{dy}{\n{g_\z (x) - y}^2}
=
C c_\s r_1 = C c_\s \delta \n{x} = C c_\s \nn{\z - \z'}_X  \n{x},
\eeas
and the proof of part (b) is complete. \prfe
We are now ready to prove part of the assertion in Proposition~\ref{frechet},
namely:

\smallskip

\noindent
{\bf Assertion 1:} 
{\em For $\g \in ]-1,1[$ and $\z \in \O$ we have $T(\g,\z) \in Y$,
and the mapping $T : ]-1,1[\times \O \to Y$ is continuous.}

\smallskip

\noindent
\prf
Let
\[
V_{\g,\z} (x) := - \int_{B_2} \frac{\r_{\g,\z}(y)}{\n{x - y}} dy,\ 
x \in \R^3,\
(\g,\z) \in ]-1,1[ \times \O.
\]
The assertions in Lemma~\ref{rho} (a) imply that
$V_{\g,\z} \in C^2(\R^3)$ with $\nabla V_{\g,\z} (0) = 0$.
Since
\[
T(\g,\z)(x) = U_0 (x) - V_{\g,\z} (g_\z (x)) - U_0 (0) + V_{\g,\z}(0),\
x \in B_3,
\]
cf.\ (\ref{tdef}), we have $T(\g,\z)(0) = 0$ and 
$T(\g,\z) \in C^1(\dot B_3) \cap C_S(B_3)$.
While we show that $T(\g,\z) \in Y$ 
for $(\g,\z) \in ]-1,1[ \times \O$ the arguments $\g$ and $\z$ remain fixed, 
and we write $V = V_{\g,\z}$. From
\[
\nabla T(\g,\z)(x) = \nabla U_0 (x) - \nabla V (g_\z (x)) Dg_\z (x),
\ x \in \dot B_3,
\]
we obtain the estimate
\[
\n{\nabla T(\g,\z)(x)} \leq \nn{D^2 U_0}_\infty \n{x} +
2 \nn{D^2 V}_\infty \n{g_\z (x)} \leq C \n{x}
\] 
with some constant $C$ which depends on $U_0$ and $V$
but not on $x$. In particular, this shows that  $T(\g,\z) \in C^1(B_3)$.
Now fix $x \in S_1$. Since any point on the line segment
$\overline{0,\, g_\z (t x)}$ can be written in the form
$g_\z (\tau x)$ with $\tau \in [0,t]$ we have
\beas
\frac{\d{x_i} T(\g,\z)(t x)}{t}
&=&
\frac{\d{x_i} U_0 (t x)}{t} - \frac{1}{t} \nabla V (g_\z (tx))
\cdot \d{x_i} g_\z (t x) \\
&=&
\frac{\d{x_i} U_0 (t x)}{t} - \frac{1}{t} 
\Bigl( D^2 V (g_\z (\tau x)) g_\z (tx)\Bigr)
\cdot \d{x_i} g_\z (t x) \\
&\to&
\nabla \d{x_i} U_0 (0) \cdot x - \left( D^2 V (0) \frac{g_\z (0 x)}{0}\right)
\cdot \d{x_i} g_\z (0x)
\eeas
as $t \searrow 0$, uniformly in $x \in S_1$, by Lemma~\ref{zeta} (d).
This shows that $T(\g,\z) \in Y$. 

To show that $T$ is continuous we fix $(\g',\z') \in ]-1,1[ \times \O$.
Constants denoted by $C$ may depend on $(\g',\z')$ but not on
$(\g,\z)\in ]-1,1[ \times \O$ or $x \in B_3$.
Restoring the subscript of $V$ we have
\beas
\nn{T(\g,\z) - T(\g',\z')}_Y 
&=&
\sup_{x \in \dot B_3} \frac{1}{\n{x}}
\Bigl| \nabla V_{\g,\z} (g_\z (x)) Dg_\z (x) -
\nabla V_{\g',\z'} (g_{\z'} (x)) Dg_{\z'} (x)\Bigr|\\
&\leq&
\sup_{x \in \dot B_3} \frac{1}{\n{x}} (I_1 + I_2 + I_3),
\eeas
where for $x \in \dot B_3$,
\beas
I_1
&:=&
\n{Dg_\z (x)}
\bigl|\nabla V_{\g,\z} (g_\z (x)) - \nabla V_{\g',\z'} (g_\z (x))\bigr| ,\\
I_2
&:=&
\n{Dg_\z (x)}
\bigl|\nabla V_{\g',\z'} (g_\z (x)) - 
\nabla V_{\g',\z'} (g_{\z'} (x))\bigr|,\\
I_3
&:=&
\bigl|\nabla V_{\g',\z'} (g_{\z'} (x))\bigr| 
\bigl|Dg_\z (x) - Dg_{\z'} (x)\bigr| .
\eeas
Using Lemma~\ref{rho} (b) and Lemma~\ref{newton} (a) with
$\s = \r_{\g,\z} - \r_{\g',\z'}$ we find
\[
\bigl|\nabla V_{\g,\z} (g_\z (x)) - \nabla V_{\g',\z'} (g_\z (x))\bigr|
\leq C\Bigl( \n{\g - \g'} + \nn{\z - \z'}_X\Bigr) \n{g_\z (x)}
\]
and thus by Lemma~\ref{defo},
\[
I_1 \leq  C \Bigl( \n{\g - \g'} + \nn{\z - \z'}_X\Bigr) \n{x},\ x \in B_3 .
\]
Since $V_{\g',\z'} \in C^2 (\R^3)$ with $\nabla V_{\g',\z'} (0) = 0$ we have
by Lemma~\ref{defo} (d),
\[
I_2 \leq C \n{g_\z (x) - g_{\z'} (x)} \leq C \nn{\z - \z'}_X \n{x},\ 
x \in B_3,
\]
and
\[
I_3 \leq \nn{D^2 V_{\g',\z'}}_\infty \n{g_{\z'} (x)}\; \nn{\z - \z'}_X
\leq C  \nn{\z - \z'}_X \n{x},\ x \in B_3,
\]
and the proof of Assertion 1 is complete. \prfe
To deal with the differentiability of $T$ we have to investigate the 
integrand
of the first term in the formula for $\d{\z} T$, 
cf.\ Proposition~\ref{frechet}: 
\begin{lemma} \label{sigma}
For $\g \in ]-1,1[$, $\z \in \O$, and $\x \in X$ define
\[
\s_{\g,\z,\x} (x) :=
\d{u} h \bigl(\g,r(x), U_\z (x)\bigr) \nabla U_\z (x) \cdot 
\frac{g_\z ^{-1} (x)}{\n{g_\z ^{-1} (x)}} \x (g_\z^{-1} (x)),\ x \in B_2,
\]
where we recall that $U_\z (x) = U_0(g_\z^{-1} (x)),\ x \in B_2$.
Then $\s_{\g,\z,\x} \in C_S (B_2)$, and there exists $C>0$ such that
for every $\g \in ]-1,1[$, $\z \in \O$, and $\x \in X$,
\[
\n{\s_{\g,\z,\x} (x)} \leq C\, \nn{\x}_X \n{x} ,\ x \in B_2 .
\]
Moreover, if we fix $(\g',\z') \in ]-1,1[ \times \O$ there exists
for each $\e >0$ a $\delta >0$ such that
for all $(\g,\z) \in ]-1,1[ \times\O$ with
$\n{\g - \g'} + \nn{\z - \z'}_X < \delta$, and $\x \in X$,
\[
\n{\s_{\g,\z,\x} (x) - \s_{\g',\z',\x} (x) } \leq \e \,\nn{\x}_X \n{x} ,
\ x \in B_2 .
\]
\end{lemma}
\prf
Since the range of $U_0$ and thus also of $U_\z$ is bounded,
the first factor in $\s_{\g,\z,\x}$ is bounded, uniformly
in $\g$ and $\z$, and the same is clearly true for the second and 
third factor.
Together with
\be \label{xigzet}
\n{\x (g_\z^{-1} (x))} \leq \nn{\x}_X \n{g_\z^{-1} (x)}
\leq 2 \nn{\x}_X \n{x},\ x \in B_2,
\ee
the estimate for $\s_{\g,\z,\x}$ follows. The continuity of
$\s_{\g,\z,\x}$ on $\dot B_2$ is clear, and at $x=0$ it follows from the
estimate above. The symmetry follows from the corresponding properties
of $U_0$, $g_\z$, and $\x$. In the following $C$ denotes a constant
which may depend on $U_0$ and $(\g',\z')$ but not on $\g,\z,\x$, or $x$.
Making excessive use of the triangle inequality we find that
\beas
&&
\Bigl|\s_{\g,\z,\x} (x) - \s_{\g',\z',\x} (x)\Bigr|\\
&&
\hspace{89pt} \leq 
C \Bigl|\d{u} h \bigl(\g,r(x), U_\z (x)\bigr) - 
\d{u} h \bigl(\g',r(x), U_{\z'} (x)\bigr)\Bigr|
\n{ \x (g_\z^{-1} (x))} \\
&&
\hspace{89pt} \quad \mbox{}+ 
C \bigl|\nabla U_\z (x) - \nabla U_{\z'} (x)\bigr| \n{ \x (g_\z^{-1} (x))} \\
&&
\hspace{89pt}  \quad \mbox{} +
C\left|\frac{g_\z ^{-1} (x)}{\n{g_\z ^{-1} (x)}}
- \frac{g_{\z'} ^{-1} (x)}{\n{g_{\z'} ^{-1} (x)}} \right|
 \n{ \x (g_\z^{-1} (x))}\\
&&
\hspace{89pt}  \quad \mbox{} +
C  \bigl| \x (g_\z^{-1} (x)) -  \x (g_{\z'}^{-1} (x))\bigr|\\
&&
\hspace{91pt} =:
I_1 + I_2 + I_3 + I_4,\ x \in \dot B_2.
\eeas
Now the estimate (\ref{xigzet})
together with the properties of the function $h$ stated in Lemma~\ref{h}
imply that
\beas
I_1 
&\leq& 
C \Bigl(\n{\g - \g'} + \n{U_0 (g_\z^{-1} (x)) - U_0(g_{\z'}^{-1} (x))}^\nu 
\Bigr) \nn{\x}_X \n{x}\\
&\leq& 
C \Bigl(\n{\g - \g'} + \n{g_\z^{-1} (x) - g_{\z'}^{-1} (x)}^\nu\Bigr) 
\nn{\x}_X \n{x}\\
&\leq&
C \Bigl(\n{\g - \g'} + \nn{\z - \z'}_X^\nu \Bigr) 
\nn{\x}_X \n{x},\ x \in B_2.
\eeas
The crucial estimate is the one for $I_2$: it is at this point
that we need the
limit condition in the definition of the Banach space $X$ and its
consequences. First note that
\beas
I_2
&\leq&
C \bigl|\nabla U_0(g_\z^{-1} (x)) - \nabla U_0 (g_{\z'}^{-1} (x))\bigr|\, 
\nn{\x}_X \n{x}\\
&&
\mbox{} + C  \bigl|Dg_\z ^{-1} (x) - Dg_{\z'}^{-1} (x)\bigr|\,
\nn{\x}_X \n{x}\\
&\leq&
C \nn{\z - \z'}_X \nn{\x}_X \n{x} +
C \n{Dg_\z ^{-1} (x) - Dg_{\z'}^{-1} (x)}\, \nn{\x}_X \n{x} .
\eeas
Now with $z:= g_\z^{-1} (x)$ and $z':= g_{\z'}^{-1} (x)$, 
\beas
\n{Dg_\z ^{-1} (x) - Dg_{\z'}^{-1} (x)} 
&=&
\n{(Dg_\z)^{-1} (z) - (Dg_{\z'})^{-1} (z')}\\
&\leq&
C \n{Dg_\z (z) - Dg_{\z'} (z')}\\
&\leq&
C \n{Dg_\z (z) - Dg_{\z'} (z)} + C \n{Dg_{\z'} (z) - Dg_{\z'} (z')} \\
&\leq&
C \nn{\z - \z'}_X + C \n{Dg_{\z'} (z) - Dg_{\z'} (z')},
\eeas
and it remains to estimate the last term in the line above.
From (\ref{dg}) we get the estimate
\beas
\n{Dg_{\z'} (z) - Dg_{\z'} (z')}
&\leq&
C\, \n{\nabla \z' (z) - \nabla \z' (z')}\\
&&
\mbox{} + C\, \n{\nabla \z' (z')}
\left| \frac{z}{\n{z}} - \frac{z'}{\n{z'}} \right|
+ \frac{C}{\n{z}} \n{\z' (z) - \z'(z')}\\
&&
\mbox{} +
C\,\n{\z' (z')} \left( \left| \frac{1}{\n{z}} - \frac{1}{\n{z'}} \right|
+ \max_{i,j =1,2,3} 
\left| \frac{z_i z_j}{\n{z}^3} - \frac{z'_i z'_j}{\n{z'}^3} \right| \right)\\
&=:&
J_1 + J_2 + J_3 + J_4 .
\eeas
With $\bar x := x /\n{x}$ there exist $s,s'>0$ such that
$z = g_\z^{-1}(x) = s \bar x$ and $z' = g_{\z'}^{-1}(x) = s' \bar x$
so that $s = \n{z}$, $s' = \n{z'}$, and
\[
\n{s - s'} = \bigl|\n{z} - \n{z'}\bigr| \leq 
\bigl|g_\z^{-1} (x) - g_{\z'} ^{-1} (x)\bigr|
\leq C \nn{\z - \z'}_X,\ x \in \dot B_2.
\]
Now given $\e >0$ we can choose $\delta >0$ according to Lemma~\ref{zeta} (c)
such that $\nn{\z - \z'}_X < \delta$ implies
\[
J_1 = C\, \n{\nabla \z' (s \bar x) - \nabla \z' (s' \bar x)} < \e,\ 
x \in \dot B_2.
\] 
Using Lemma~\ref{defo} (d) and Lemma~\ref{zeta} we obtain
\[
J_2 \leq C \left( \frac{1}{\n{z}} + \frac{1}{\n{z'}} \right) \n{z - z'}
\leq \frac{C}{\n{x}} \nn{\z - \z'}_X \n{x}
\leq C \nn{\z - \z'}_X,
\]
\[
J_3 \leq \frac{C}{\n{z}} \n{z - z'} \leq C \nn{\z -\z'}_X,
\]
and
\[
J_4 \leq C \n{z'} \left( \frac{1}{\n{z}^2} + \frac{1}{\n{z'}^2} \right)
\n{z - z'} \leq C \nn{\z - \z'}_X, 
\]
so that finally
\[
I_2 \leq C \nn{\z - \z'}_X \nn{\x}_X \n{x} + C\, \e\,\nn{\x}_X \n{x},
\ x \in \dot B_2,
\]
provided $\nn{\z - \z'}_X < \delta$.
The remaining terms $I_3$ and $I_4$ are much easier to estimate:
\beas
I_3 
&\leq&
C  \left( \frac{1}{\n{g_\z^{-1} (x)}} +
\frac{1}{\n{g_{\z'}^{-1} (x)}}\right) 
\bigl|g_\z^{-1} (x) - g_{\z'}^{-1} (x)\bigr|
\nn{\x}_X \n{x} \\
&\leq&
C \nn{\z - \z'}_X \nn{\x}_X \n{x},\ x \in \dot B_2,
\eeas
and
\[
I_4 \leq C \nn{\x}_X \bigl|g_\z ^{-1} (x) - g_{\z'}^{-1} (x)\bigr| 
\leq C \nn{\x}_X \nn{\z - \z'}_X \n{x},\ x \in \dot B_2,
\]
and the proof of Lemma~\ref{sigma} is complete. \prfe
To continue with the proof of Proposition~\ref{frechet}
we denote for fixed $(\g,\z) \in ]-1,1[\times \O$ 
by $L\x$ the right hand side of the definition of $\d{\z} T(\g,\z)\x$,
$\x \in X$. We now show:

\smallskip

\noindent
{\bf Assertion 2:}
{\em $L \in {\cal L}(X,Y)$ is a bounded, linear operator, and
for all $\x \in X$, 
\[
\lim_{t \to 0} \frac{T(\g,\z + t \x) - T(\g,\z)}{t} = L \x
\]
with respect to $\nn{\cdot}_Y$.}

\smallskip

\noindent
\prf
It is convenient to introduce the auxiliary Banach space
\beas
\overline{Y} := \Bigl\{ f \in C_S (B_3) 
&\mid& 
f(0) = 0,\ f \in C^1 (B_3),\\
&&
\exists C>0 : \n{\nabla f (x)} \leq C \n{x},\ x \in B_3\Bigr\},
\eeas
which we equip with the norm $\nn{\cdot}_Y$; clearly, $Y$ is a closed
subspace of $\overline{Y}$. Since we already know that $T$ maps $X$ into $Y$
it is then sufficient to show that $L \in {\cal L}(X,\overline{Y})$ 
and that the asserted convergence holds.  
To see the former define
\[
V_\x (x) := - \int_{B_2} \frac{1}{\n{x - y}} 
\s_{\g,\z,\x}(y)\, dy,\ x \in \R^3,
\]
and
\[
W (x) := -  \int_{B_2} \frac{1}{\n{x - y}} \r_{\g,\z}(y)\, dy,\ x \in \R^3 .
\]
Then $V_\x \in C^1 (\R^3)$, $W \in C^2 (\R^3)$, and we can write
\[
(L\x)(x) = V_\x (g_\z (x)) - V_\x (0) - \nabla W (g_\z (x)) \cdot 
\frac{x}{\n{x}} \x (x),\ x \in B_3.
\]
This implies that for $\x \in X$ we have $L\x \in C^1(\dot B_3)$,
$(L\x)(0) = 0$, and
\beas
(\nabla L\x)(x)
&=&
\nabla V_\x (g_\z (x)) D g_\z (x) - D^2 W (g_\z (x)) D g_\z (x) 
\frac{x}{\n{x}} \x (x)\\
&&
- \nabla W (g_\z (x)) D \left(\frac{x}{\n{x}}\right) \x (x)
- \nabla W (g_\z (x)) \cdot \frac{x}{\n{x}} \nabla \x (x),\ x \in \dot B_3.
\eeas
Using Lemma~\ref{sigma} and Lemma~\ref{newton} (a) we obtain the estimate
\beas
\n{(\nabla L\x)(x)}
&\leq&
C \nn{\x}_X \n{g_\z (x)} + C \n{\x (x)} + C \nn{D^2 W}_\infty \n{g_\z (x)}\, 
\left( \frac{\n{\x (x)}}{\n{x}} + \n{\nabla \x (x)} \right) \\
&\leq&
C \nn{\x}_X \n{x},\ x \in \dot B_3 .
\eeas
In particular, this implies that $L\x$ is differentiable also at $x=0$,
and
\[
\nn{L \x}_Y \leq C\,\nn{\x}_X,\ \x \in X .
\]
The symmetry of $L\x$ follows easily from the corresponding properties
of $V_\x$, $W$, $\z$, and $\x$. In order to show that $L\x$ is indeed
the Gateaux derivative of $T$ at $(\g,\z)$ in the direction of $\x$
we choose $t_0 >0$ such that $\z + t \x \in \O$ for $\n{t} < t_0$.
Although this is in conflict with earlier notation it is convenient
to abbreviate
\[
g_t (x) = g_{\z + t \x} (x) = x + 
\bigl(\z (x) + t \x (x)\bigr) \frac{x}{\n{x}},\
x \in B_3,\ t \in ]-t_0,t_0[ .
\]
Then for each $x \in \dot B_2$ the mapping
$]-t_0,t_0[ \ni t \mapsto g_t^{-1} (x)$
is continuously differentiable with
\[
\frac{d}{dt}g_t^{-1} (x) = - (Dg_t)^{-1} (g_t^{-1} (x)) \x(g_t^{-1} (x))
\frac{g_t^{-1} (x)}{\n{g_t^{-1} (x)}}.
\]
To see this, define for fixed $z \in \dot B_2$ the mapping
\[
G(t,x) := g_t (x) - z,\ t \in ]-t_0,t_0[,\ x \in \dot B_3 .
\]
Since $G(t,g_t^{-1} (z)) = 0,\ t \in ]-t_0,t_0[$, the asserted regularity
of $g_t^{-1}$ with respect to $t$ follows from the regularity of $G$,
the fact that $\d{x} G(t,x) = D g_t (x)$ is invertible, and the implicit 
function theorem. If we now differentiate the identity 
$x = g_t(g_t^{-1} (x))$
with respect to $t$ we obtain the formula for $\frac{d}{dt}g_t^{-1} (x)$.

It will also be convenient to abbreviate
\[
\r_t (x) := \r_{\g,\z + t \x} (x),\ \s_t (x) :=\s_{\g,\z+t \x,\x} (x),\ 
t \in ]-t_0,t_0[,\ x \in B_2,
\]
and define
\[
F(t,x) := \int_{B_2} \left(\frac{1}{\n{x - y}} - \frac{1}{\n{y}}\right)
\r_t (y)\, dy,\ x \in \R^3,\ t \in ]-t_0,t_0[ .
\]
Then except for $\d{t}^2 F$ all derivatives of $F$ up to second
order exist and are continuous on $]-t_0,t_0[ \times \R^3$, and 
\beas
\d{t} F(t,x) 
&=&
- \int_{B_2} \left(\frac{1}{\n{x - y}} - \frac{1}{\n{y}}\right)
\s_t (y)\, dy,\\
\nabla F(t,x) 
&=&
-  \int_{B_2} \frac{x-y}{\n{x-y}^3} \r_t (y)\, dy .
\eeas
These results follow easily from the fact that $\r_t \in C^1_c (\open{B}_2)$
and 
\beas
\frac{d}{dt} \r_t (y) 
&=&
\d{u} h\bigl(\g,r(y),U_0(g_t^{-1}(y))\bigr)  \nabla U_0 (g_t^{-1} (y))
\cdot \frac{d}{dt} g_t^{-1} (x)\\
&=&
-\d{u} h\bigl(\g,r(y),U_0(g_t^{-1}(y))\bigr)  \nabla U_0 (g_t^{-1} (y))
(Dg_t)^{-1} (g_t^{-1}(y)) \\
&&\quad
\x (g_t^{-1} (y))  \frac{g_t^{-1} (y)}{\n{g_t^{-1} (y)}}\\
&=&
-\d{u} h\bigl(\g,r(y),U_0(g_t^{-1}(y))\bigr)  
\nabla \bigl(U_0 (g_t^{-1} (y))\bigr)
\cdot \frac{g_t^{-1} (y)}{\n{g_t^{-1} (y)}} \x (g_t^{-1} (y))\\
&=&
- \s_t (y).
\eeas
Now
\beas
\frac{T(\g,\z + t \x)(x) - T(\g,\z)(x)}{t} 
&=&
\frac{F(t,g_t (x)) - F(0,g_t(x))}{t}\\
&&
\mbox{} +
\frac{F(0,g_t (x)) - F(0,g_0 (x))}{t},\ t \in ]-t_0,t_0[,\ x \in B_3,
\eeas
and
\[
(L\x)(x) = \d{t} F(0,g_0 (x)) 
+ \nabla F (0,g_0 (x)) \cdot \frac{x}{\n{x}} \x (x),
\ x \in B_3;
\]
one should be careful to note that here
$g_0 = g_{\z + 0 \x} = g_\z$. We claim that as $t \to 0$,
\be \label{gat1}
\frac{F(t,g_t (x)) - F(0,g_t(x))}{t} \to  \d{t} F(0,g_0 (x))
\ee
and
\be \label{gat2}
\frac{F(0,g_t (x)) - F(0,g_0 (x))}{t} \to \nabla F (0,g_0 (x)) \cdot 
\frac{x}{\n{x}} \x (x),
\ee
where both limits are understood with respect to the norm $\nn{\cdot}_Y$.
This would then prove that $L$ is the Gateaux differential of
$T$ at $(\g,\z)$. As to (\ref{gat1}) we observe that
\beas
&&
\left| \nabla \frac{F(t,g_t (x)) - F(0,g_t(x))}{t} 
- \nabla\Bigl( \d{t} F(0,g_0 (x))\Bigr) \right|\\
&&
\hspace{65pt} =
\left| \frac{\nabla F(t,g_t (x)) - \nabla F(0,g_t(x))}{t} Dg_t (x) 
- \nabla \d{t} F(0,g_0 (x)) Dg_0 (x) \right|\\
&&
\hspace{65pt}\leq
\left| \frac{\nabla F(t,g_t (x)) - \nabla F(0,g_t(x))}{t}  
- \nabla \d{t} F(0,g_t (x)) \right| \n{Dg_t (x)}\\
&&
\hspace{65pt}\quad \mbox{} +
\bigl| \nabla\d{t}  F(0,g_t (x)) - \nabla\d{t}  F(0,g_0(x))\bigr|
\n{Dg_t (x)}\\
&&
\hspace{65pt}\quad \mbox{} + 
\bigl| \nabla \d{t}F(0,g_0 (x))\bigr| \bigl|D g_t (x) - Dg_0 (x)\bigr|\\
&&
\hspace{65pt} 
=: I_1 + I_2  + I_3 .
\eeas
Let $\e >0$. For every $z \in \R^3$ there exists $\tau$ between
$0$ and $t$ such that 
\beas
\left| \frac{\nabla F(t,z) - \nabla F(0,z)}{t}  
- \nabla \d{t} F(0,z) \right|
&=&
\left|  \nabla \d{t} F(\tau,z) - \nabla \d{t} F(0,z) \right|\\
&=&
\left| \nabla \int_{B_2} 
\frac{1}{\n{z - y}} (\s_\tau (y) - \s_0 (y))\, dy \right|
\eeas
and using Lemma~\ref{newton} (a), the latter integral can be estimated by 
$C\, \e\, \nn{\x}_X \n{z}$, provided 
\[
\n{\s_\tau (y) - \s_0 (y)} \leq \e\, \nn{\x}_X \n{y},\ y \in B_2.
\]
Lemma~\ref{sigma} therefore implies that for $\delta >0$ sufficiently
small we have
\[
I_1 \leq C \e \n{g_t (x)} \leq C \,\e \,\n{x},\ x \in B_3,
\]
provided $\n{t} < \delta$. Note that $C$ depends on $\z$ and $\x$,
but not on $t$ or $x$.
Again by Lemma~\ref{sigma}
and Lemma~\ref{newton} (b) we find the estimate
\[
I_2 \leq 
C \,\nn{\x}_X \nn{\z + t \x - \z}_X^{1/2} \n{x} \leq C \,\n{t}^{1/2} \n{x},
\ x \in B_3,
\]
and by Lemma~\ref{sigma}, Lemma~\ref{newton} (a), and Lemma~\ref{defo} (d) 
we conclude that
\[
I_3 \leq
C\, \nn{\x}_X \n{g_0 (x)}\, \nn{\z + t \x - \z}_X 
\leq C \,\n{t}\, \n{x},\ x \in B_3.
\] 
This proves the convergence in (\ref{gat1}) with respect to $\nn{\cdot}_Y$.
As to (\ref{gat2}) we observe that for every $x \in B_3$,
\[
\frac{F(0,g_t (x)) - F(0,g_0 (x))}{t} =
\frac{d}{dt} F(0,g_t (x))_{|t = \tau} =
\nabla F(0,g_\tau (x)) \cdot \frac{x}{\n{x}} \x (x)
\]
where $\tau$ lies between $0$ and $t \in ]-t_0,t_0[$. Therefore,
\beas
&&
\left|\nabla \frac{F(0,g_t (x)) - F(0,g_0 (x))}{t} -
\nabla\Bigl( \nabla F (0,g_0 (x)) 
\cdot \frac{x}{\n{x}} \xi (x) \Bigr) \right|\\
&&
\hspace{94pt}=
\left|\nabla \Bigl[ 
\Bigl( \nabla F (0,g_\tau (x)) - \nabla F (0,g_0 (x)) \Bigr) \cdot 
\frac{x}{\n{x}} \xi (x) \Bigr] \right|\\
&&
\hspace{94pt}\leq
\bigl| D^2 F(0,g_\tau (x)) - D^2 F(0,g_0(x))\bigr|\, C \,\n{x} \\
&&
\hspace{94pt}\quad \mbox{} +
\n{D^2 F(0,g_0(x))} \, \bigl|D g_\tau (x) - D g_0 (x)\bigr|\, C \,\n{x} \\
&&
\hspace{94pt}\quad \mbox{} +
\bigl| \nabla F (0,g_\tau (x)) - \nabla F(0,g_0 (x))\bigr| 
\left| D \left( \x (x) \frac{x}{\n{x}}\right)\right|\\
&&
\hspace{94pt}\leq
C \,\n{t}\, \n{x} + C \,\bigl| D^2 F(0,g_\tau (x)) - D^2 F(0,g_0(x))\bigr|\, 
\n{x},\
x \in B_3 .
\eeas
Since $D^2 F(0,\cdot)$ is uniformly continuous on $B_4$, which contains
$g_\tau (x)$ for $x \in B_3$ and $\tau \in ]-t_0,t_0[$,
cf.\ Lemma~\ref{defo} (b), and 
\[
\n{g_\tau (x) - g_0 (x)} \leq \nn{\x}_X \n{\tau} \leq C \n{t},\ x \in B_3,
\]
we obtain the convergence in (\ref{gat2}) with respect to the norm
$\nn{\cdot}_Y$. This completes the proof of Assertion 2. \prfe
Since a continuous Gateaux derivative is a Fr\'{e}chet derivative
the proof of Proposition~\ref{frechet} will be complete, once we show:

\smallskip

\noindent
{\bf Assertion 3:}
{\em The mapping $]-1,1[ \times \O \ni (\g,\z) \to \d{\z} T(\g,\z)
\in {\cal L}(X,Y)$ is continuous.}

\smallskip

\noindent
\prf
Fix $(\g',\z') \in ]-1,1[ \times \O$ and let 
$(\g,\z) \in ]-1,1[ \times \O$  and $\x \in X$ with $\nn{\x}_X =1$.
Since
\beas
&&
\bigl[\d{\z} T(\g,\z)\x\bigr](x) - \bigl[\d{\z} T(\g',\z') \x\bigr] (x)\\
&&
\qquad =
-  \int_{B_2} \left[\left( \frac{1}{\n{g_\z (x) - y}}-\frac{1}{\n{y}} \right)
 \s_{\g,\z,\x} (y) 
- \left( \frac{1}{\n{g_{\z'} (x) - y}} - \frac{1}{\n{y}} \right)
 \s_{\g',\z',\x} (y) \right]\, dy\\
&&
\qquad \quad \mbox{} -
\int_{B_2} \left[ \frac{g_\z (x) - y}{\n{g_\z (x) - y}^3} \r_{\g,\z}(y)
- \frac{g_{\z'} (x) - y}{\n{g_{\z'} (x) - y}^3} \r_{\g',\z'}(y)
\right] dy \cdot \frac{x}{\n{x}} \x (x),\ x \in B_3,
\eeas
we find
\[
\Bigl| \nabla\Bigl(\bigl[\d{\z} T(\g,\z)\x\bigr](x) - 
\bigl[\d{\z} T(\g',\z') \x\bigr] (x)
\Bigr) \Bigr| \leq I_1 + I_2 + I_3 + I_4 + I_5 + I_6,
\]
where
\beas
I_1
&:=&
\left| \nabla \int_{B_2}
\left( \frac{1}{\n{g_\z (x) - y}} - \frac{1}{\n{y}} \right)
\Bigl( \s_{\g,\z,\x}(y) - \s_{\g',\z',\x}(y) \Bigr) dy \right|,\\
I_2
&:=&
\left| \nabla \int_{B_2}
\left( \frac{1}{\n{g_\z (x) - y}} - \frac{1}{\n{g_{\z'} (x) - y}} \right)
\s_{\g',\z',\x}(y)  dy \right|,\\
I_3
&:=&
\left| D \int_{B_2} \frac{g_\z (x) - y}{\n{g_\z (x) - y}^3} 
\bigl(\r_{\g,\z}(y) - \r_{\g',\z'}(y) \bigr) \, dy\right| \n{\x (x)},\\
I_4
&:=&
\left| D \int_{B_2}  
\left( \frac{g_\z (x) - y}{\n{g_\z (x) - y}^3}
- \frac{g_{\z'} (x) - y}{\n{g_{\z'} (x) - y}^3} \right)  \r_{\g',\z'}(y)\,
dy\right| \n{\x (x)},\\
I_5
&:=&
\left| \int_{B_2}  
\frac{g_\z (x) - y}{\n{g_\z (x) - y}^3}
\bigl( \r_{\g,\z}(y) -  \r_{\g',\z'}(y)\bigr)\, dy \right| 
\left|D\left(\x (x) \frac{x}{\n{x}}\right)\right|,\\
I_6
&:=&
\left| \int_{B_2}  
\left( \frac{g_\z (x) - y}{\n{g_\z (x) - y}^3} -
\frac{g_{\z'} (x) - y}{\n{g_{\z'} (x) - y}^3} \right) \r_{\g',\z'}(y) 
\, dy \right|
\left|D\left(\x (x) \frac{x}{\n{x}}\right)\right| .
\eeas
Given $\e >0$ we choose $\delta>0$ so that the second estimate 
in Lemma~\ref{sigma}
holds for $\n{\g - \g'} + \nn{\z - \z'}_X < \delta$. Then
Lemma~\ref{newton} (a) implies, with $z = g_\z (x)$, the estimate
\[
I_1 \leq
\left| \nabla \int_{B_2}
\frac{1}{\n{z - y}}
\Bigl( \s_{\g,\z,\x}(y) - \s_{\g',\z',\x}(y) \Bigr) dy
D g_\z (x) \right| \leq
C\, \e\, \n{z} \leq C\, \e\, \n{x},\ x \in B_3 .
\]
Defining
\[
V(x) := \int_{B_2} \frac{1}{\n{x-y}} \s_{\g',\z',\x} (y)\, dy,\ x \in \R^3,
\]
we have $V \in C^1 (\R^3)$ with $\nabla V (0) = 0$, and
\beas
I_2
&=&
\Bigl| \nabla V (g_\z (x)) D g_\z (x) - 
\nabla V (g_{\z'} (x)) D g_{\z'} (x)\Bigr| \\
&\leq&
C \Bigl| \nabla V (g_\z (x))  -  \nabla V (g_{\z'} (x)) \Bigr|
+ \n{\nabla V (g_{\z'} (x)} \n{D g_\z (x) - D g_{\z'} (x)} \\
&\leq&
C \nn{\z - \z'}_X^{1/2} \n{x} + C \n{g_{\z'}(x)} \nn{\z - \z'}_X 
\leq
C \nn{\z - \z'}_X^{1/2} \n{x},\ x \in B_3,
\eeas
where we have used the first estimate in Lemma~\ref{sigma}, 
Lemma \ref{newton}, and Lemma~\ref{defo}. 
In order to estimate the remaining terms we define
\[
V_{\g,\z} (x) := \int_{B_2} \frac{1}{\n{x - y}} \r_{\g,\z} (y)\, dy,\
x \in \R^3 .
\]
Then $V_{\g,\z} \in C^2 (\R^3)$ and
\beas
I_3
&=&
\Bigl| D^2 V_{\g,\z} (g_\z (x)) - D^2 V_{\g',\z'} (g_\z (x)) \Bigr| 
\n{D g_\z (x)}\, \n{\x (x)} \\
&\leq&
C \n{x}\, \nn{\r_{\g,\z} - \r_{\g',\z'}}_1^{1/16}
 \nn{\r_{\g,\z} - \r_{\g',\z'}}_\infty^{3/4}
 \nn{\nabla \r_{\g,\z} - \nabla \r_{\g',\z'}}_\infty^{3/16}\\
&\leq&
C \Bigl(\n{\g - \g'} + \nn{\z - \z'}_X\Bigr)^{3/4} \n{x},\ x \in B_3,
\eeas
where we have used Lemma~\ref{rho} and \cite[Lemma 1]{BD}. Next we have
\beas
I_4
&=&
\Bigl| D^2 V_{\g',\z'} (g_\z (x)) Dg_\z (x) - 
D^2 V_{\g',\z'}(g_{\z'}(x)) Dg_{\z'} (x) \Bigr|\, \n{\x (x)} \\
&\leq&
C\, \n{x}\, 
\Bigl|D^2 V_{\g',\z'} (g_\z (x))  -  D^2 V_{\g',\z'}(g_{\z'}(x) \Bigr|
+ C\, \n{x}\, \n{Dg_\z (x) - Dg_{\z'} (x)} \\
&\leq&
\e\, \n{x} + C \, \n{x} \,\nn{\z - \z'}_X,\ x \in B_3,
\eeas
provided $\nn{\z - \z'}_X$ is small enough, where we have used the fact that
$D^2 V_{\g',\z'}$ is uniformly continuous on $B_4 \ni g_\z (x),g_{\z'} (x)$
and $\n{g_\z (x) - g_{\z'} (x)} \leq \nn{\z - \z'}_X$. 
By Lemma~\ref{rho} (b) and Lemma~\ref{newton} (a) for
$\s = \r_{\g,\z} - \r_{\g',\z'}$ we obtain, with $z = g_\z (x)$,
\[
I_5 \leq
C \left| \nabla \int_{B_2} \frac{1}{\n{z-y}} 
\Bigl(\r_{\g,\z}(y) - \r_{\g',\z'} (y)\Bigr)\, dy \right| \leq
C \Bigl(\n{\g - \g'} + \nn{\z - \z'}\Bigr) \n{x},\ x \in B_3.
\]
Finally,
\beas
I_6
&\leq&
C \Bigl|\nabla V_{\g',\z'} (g_\z (x)) - \nabla V_{\g',\z'} (g_{\z'}(x))\Bigr|
\leq
C \nn{D^2 V_{\g',\z'}}_\infty \n{g_\z (x) - g_{\z'} (x)}\\
&\leq&
C \nn{\z - \z'}_X \n{x},\ x \in B_3.
\eeas
We have shown that for every $\e>0$ there exists
$\delta >0$, depending on $(\g',\z')$, such that for all
$(\g,\z) \in ]-1,1[\times \O$ with $\n{\g - \g'} + \nn{\z - \z'}_X
< \delta$ and all $\x \in X$ with $\nn{\x}_X = 1$ we have
\[
\nn{ \d{\z} T(\g,\z)\x - \d{\z} T(\g',\z')\x}_Y \leq \e,
\]
and the proof of Proposition~\ref{frechet} is complete. \prfe 
\section{$\dz T(0,0)$ is an isomorphism}
\setcounter{equation}{0}
The aim of this section is to prove the following result:
\begin{prop} \label{isom} 
The mapping $\d{\z} T(0,0) : X \to Y$ is a linear isomorphism.
\end{prop}
Let us abbreviate $L_0 \x := \d{\z} T(0,0)\x$ for $\x \in X$.
In order to prove the result above we rewrite $L_0 \x$:
Observe first that $g_0 = id$, and therefore $U_\z$, defined in 
Proposition~\ref{frechet},
coincides with the potential $U_0$ of the spherically symmetric 
steady state we started with, if $\z = 0$. In particular,
$\r_0 (\n{x}) = h_0 (U_0(\n{x})) = h(0,r(x),U_0 (\n{x})$ for $x \in \R^3$, 
and
\[
\r_0'(\n{x}) = \d{u} h\bigl(0,r(x), U_0 (\n{x})\bigr) U_0'(\n{x})
= \d{u} h\bigl(0,r(x), U_0(x)\bigr) \nabla U_0 (x)\cdot \frac{x}{\n{x}},\ 
x \in \R^3.
\]   
Therefore,
\beas
(L_0 \x)(x) 
&=&
- \int_{B_2} \left( \frac{1}{\n{x - y}} - \frac{1}{\n{y}} \right) 
\r_0' (\n{y}) \x (y)\, dy  - \int_{B_2} \frac{x - y}{\n{x - y}^3}
\r_0 (\n{y}) \, dy \cdot \frac{x}{\n{x}} \x (x)\\
&=&
- U_0'(\n{x}) \x (x) -\int_{B_2} 
\left( \frac{1}{\n{x - y}} - \frac{1}{\n{y}} \right) 
\r_0' (\n{y}) \x (y)\, dy,\ x \in B_3 ,\ \x \in X.
\eeas
Now let
\[
(K\x)(x) := - \frac{1}{U_0'(\n{x})} 
\int_{B_2} \left( \frac{1}{\n{x - y}} - \frac{1}{\n{y}} \right) 
\r_0' (\n{y}) \x (y)\, dy,\ x \in B_3,\ \x \in C_S (B_3) .
\]
Then we can write
\be \label{lo}
(L_0 \x)(x) = - U_0'(\n{x}) \bigl[(id - K)\x\bigr](x),\ x \in B_3,\ \x \in X.
\ee
As a first step towards proving Proposition~\ref{isom} we show:

\smallskip

\noindent
{\bf Assertion 1:}
{\em The linear operator $K : C_S(B_3) \to C_S(B_3)$ is compact, where
$C_S(B_3)$ is equipped with the supremum norm $\nn{\cdot}_\infty$.}

\smallskip

\noindent 
\prf
For $\x \in C_S(B_3)$ let
\be \label{v}
V_\x (x) := - \int_{B_2} \frac{1}{\n{x-y}} \r_0' (\n{y}) \x (y)\, dy,
\ x \in \R^3.
\ee
Then $V_\x \in C^1 (\R^3)$, $\nabla V_\x (0) = 0$, and 
\[
(K\x)(x) = \frac{1}{U_0'(\n{x})} \bigl( V_\x (x) - V_\x (0) \bigr),\ 
x \in B_3 .
\]
Using Lemma~\ref{ss} (c) we obtain the estimate
\[
\n{(K\x)(x)} \leq \frac{1}{C\n{x}}\nn{\nabla V_\x }_\infty \n{x}
\leq C \nn{\x}_\infty,\ x \in B_3,
\] 
where the constant $C$ depends on $\r_0$ and $U_0$, but not on $\x$ or
$x$. Thus $K$ maps bounded sets into bounded sets.
We claim that $K\x$ is H\"older continuous
with exponent $1/2$, uniformly on bounded sets in $C_S(B_3)$.
Let $M>0$ and assume $\nn{\x}_\infty \leq M$.
In the following, constants denoted by $C$ depend on $\r_0$,
$U_0$, and $M$, but not on $\x$ itself. 
There exists a constant $C>0$ such that
\be \label{nphoe}
\n{\nabla V_\x (x) - \nabla V_\x (x')} \leq C \nn{\r_0' \x}_\infty 
\n{x - x'}^{1/2},\ x,x' \in B_3,
\ee
cf.~\cite[Probl.~4.8]{GT}. Since $\nabla V_\x (0) = 0$, (\ref{nphoe}) implies
\[
\n{\nabla V_\x (x)} \leq C \n{x}^{1/2},\ x \in B_3.
\]
Now let $x,x' \in \dot B_3$ and $\n{x} \leq \n{x'}$. Then
\beas
\n{(K\x)(x) - (K\x)(x')} 
&\leq& 
\left| \frac{1}{U_0'(\n{x})} - \frac{1}{U_0'(\n{x'})}\right|
\n{ V_\x (x) - V_\x (0) }\\
&&
\mbox{} + \frac{1}{U_0'(\n{x'})} \n{V_\x (x) - V_\x (x')} =: I_1 + I_2 .
\eeas
and we obtain for some $z \in B_3$ with $\n{z} \leq \n{x'}$
the estimates
\beas
I_1
&\leq&
C \frac{\n{U_0'(\n{x}) - U_0'(\n{x'})}}{\n{x} \n{x'}}
\n{\nabla V_\x (z)} \n{x} \leq
C \n{x - x'}^{1/2} \frac{(\n{x} + \n{x'})^{1/2}}{\n{x'}} \n{z}^{1/2}\\
&\leq&
C \n{x - x'}^{1/2},
\eeas
and
\[
I_2 \leq
\frac{C}{\n{x'}} \n{\nabla V_\x (z)}\n{x - x'} \leq
\frac{C}{\n{x'}} \n{z}^{1/2}\n{x - x'}  \leq
C \n{x - x'}^{1/2} 
\]
so that
\[
\bigl| (K\x)(x) - (K\x)(x')\bigr| \leq C \n{x - x'}^{1/2},\ 
x, x' \in \dot B_3.
\]
Also,
\[
\n{(K\x)(x)} \leq C \n{\nabla V_\x (z)} \leq C \n{x}^{1/2},\ x \in \dot B_3,
\]
and we have shown that $K$ maps bounded subsets of $C_S(B_3)$
into bounded and equicontinuous subsets of $C_S(B_3)$.
Thus $K$ is compact by the Arzela-Ascoli theorem,
and the proof of Assertion 1 is complete. \prfe
As second step in the proof of Proposition~\ref{isom} we show:

\smallskip

\noindent
{\bf Assertion 2:}
{\em $id - K : C_S(B_3) \to C_S(B_3)$ is one-to-one and onto.}

\smallskip

\noindent 
\prf
Since $K$ is compact is suffices to show that $id - K$ is one-to-one.
Let $\x \in C_S(B_3)$ with $\x - K\x =0$. In order to show that $\x = 0$
we expand $\x$ into spherical harmonics $Y_{lm}$, $l \in \N_0,\
m=-l,\ldots,l$, where we use the notation of \cite[Ch.~3]{J} concerning the
latter.
Denote by $(r,\theta,\phi)$ and
$(s,\tau,\psi)$ the polar coordinates of a point
$x$ or $y \in B_3$ respectively. For $l \in \N_0$ and
$m = -l,\ldots,l$ we define
\[
\xi_{lm} (r) := \frac{1}{r^2} \int_{\n{x}=r} Y_{lm}^\ast (\theta,\phi)
\xi (x)\,dS_x .
\] 
Using the expansion 
\[
\frac{1}{\n{x-y}} =
\sum_{l=0}^\infty \sum_{m=-l}^l \frac{4 \pi}{2 l + 1} \frac{r_<^l}{r_>^{l+1}}
Y_{lm}^\ast (\tau,\psi) Y_{lm}(\theta,\phi),
\]
where $r_< := \min(r,s)$ and $r_> := \max(r,s)$,
cf.\ \cite[Eqn.~(3.70)]{J}, we find that
\beas
\xi_{lm} (r)
&=&
- \frac{1}{r^2 U_0'(r)} \int_{B_3} 
\int_{\n{x}=r} \left( \frac{1}{\n{x-y}} - \frac{1}{\n{y}}\right)
Y_{lm}^\ast (\theta,\phi)\,dS_x \r_0'(s) \xi (y)\, dy\\ 
&=&
- \frac{4 \pi}{2 l + 1} \frac{1}{U_0'(r)}
\int_0^3 \r_0'(s) \left( \frac{r_<^l}{r_>^{l+1}} - \frac{0^l}{s^{l+1}}\right)
\int_{\n{y}=s} Y_{lm}^\ast (\tau,\psi)\x (y)\, dS_y\, ds \\
&=&
- \frac{4 \pi}{2 l + 1} \frac{1}{U_0'(r)}
\int_0^3 \r_0'(s) \left( \frac{r_<^l}{r_>^{l+1}} - \frac{0^l}{s^{l+1}}\right)
s^2 \xi_{lm} (s)\, ds,\ r \in ]0,3].
\eeas
This implies that
\[
\x_{00} (r) = - \frac{4 \pi}{r U_0'(r)} \int_0^r \r_0'(s)\, s(s-r)\, 
\xi_{00}(s)\, ds .
\]
Clearly, $\xi_{00}$ vanishes in the limit at $r=0$. Let $R\geq 0$ be such
that $\x_{00}$ vanishes on $[0,R]$. Then for $r \in [R,3]$,
\[
\n{\x_{00}(r)} 
\leq
\frac{4 \pi}{r U_0'(r)} \nn{\r_0'}_\infty
\sup_{0 \leq s \leq r} \n{\x_{00}(s)} \int_R^r s(r-s)\, ds
\leq
C\, (r-R)\, \sup_{0 \leq s \leq r} \n{\x_{00}(s)}.
\]
This implies that $\x_{00}$ vanishes in a right neighborhood of $R$
and thus on the whole interval $[0,3]$. Up to multiplicative constants the 
spherical harmonics for $l=1$ are given by
$\sin \theta e^{\pm i \phi}$ and $\cos \theta$, and the fact that
$\x \in C_S$ implies that $\xi_{1-1} = \xi_{10} = \xi_{11} = 0$.
Let $l \geq 2$. Then
\[
\xi_{lm} (r)
= - \frac{4 \pi}{2 l + 1} \frac{1}{U_0'(r)}
\left( \int_0^r \r_0'(s) \frac{s^l}{r^{l+1}} s^2 \xi_{lm} (s)\, ds
      + \int_r^3 \r_0'(s) \frac{r^l}{s^{l+1}} s^2 \xi_{lm} (s)\, ds \right),
\]
and
\beas
\n{\xi_{lm} (r)}
&\leq&
\frac{4 \pi}{2 l + 1} \frac{1}{U_0'(r)} \nn{\x_{lm}}_\infty
\left(\frac{1}{r^2} \int_0^r (- \r_0')(s) \frac{s^{l-1}}{r^{l-1}} s^3\, ds
+ r \int_r^3 (- \r_0')(s) \frac{r^{l-1}}{s^{l-1}}\, ds \right)\\
&\leq&
\frac{4 \pi}{2 l + 1} \frac{1}{U_0'(r)} \nn{\x_{lm}}_\infty
\left(\frac{1}{r^2} \int_0^r (- \r_0')(s) s^3\, ds
+ r \int_r^3 (- \r_0')(s)\, ds \right)\\
&=&
\frac{4 \pi}{2 l + 1} \frac{1}{U_0'(r)} \nn{\x_{lm}}_\infty
\left(- \frac{1}{r^2} r^3 \r_0(r) + \frac{3}{r^2} \int_0^r \r_0(s) s^2\, ds
+ r \r_0 (r)\, ds \right)\\
&=&
\frac{3}{2l+1} \nn{\x_{lm}}_\infty,
\eeas
and since $2 l + 1 > 3$ this implies that $\xi_{lm}$ vanishes for $l \geq 2$
as well.
We have shown that $id - K$ is one-to-one as claimed, and 
Assertion 2 is therefore established. \prfe
It is clear that $L_0 : X \to Y$ is now one-to-one as well: just 
observe (\ref{lo}) and the fact that 
$U_0'(r) >0$ for $r>0$. It remains to show:

\smallskip

\noindent
{\bf Assertion 3:} {\em $L_0 : X \to Y$ is onto.}

\smallskip

\noindent
\prf
Let $g \in Y$ and define $q := g/U_0'$. We claim that $q \in X$.
To see this we first observe that $q \in C^1(\dot B_3) \cap C_S (B_3)$,
and 
\[
\n{\nabla q(x)}
\leq
\frac{\n{\nabla g (x)}}{U_0'(\n{x})} + 
\n{g(x)} \left| \frac{U_0''(\n{x})}{U_0'(\n{x})^2} \frac{x}{\n{x}} \right|
\leq
C \left( \frac{\n{\nabla g (x)}}{\n{x}} +  \frac{\n{ g (x)}}{\n{x}^2}\right)
\leq 2 C \, \nn{g}_Y.  
\]  
By definition of $Y$ and since $U_0 \in C^2([0,\infty[)$ with $U_0''(0) > 0$
we have that for every $x \in S_1$,
\beas
\nabla q (tx)
&=&
\frac{\nabla g (t x)}{t} \frac{t}{U_0'(t)} -
\frac{g(t x)}{t^2} U_0''(t) \left(\frac{t}{U_0'(t)}\right)^2 x\\
&\to&
\frac{\nabla g (0 x)}{0} \frac{1}{U_0''(0)} -
\frac{g(0 x)}{0^2} U_0''(0) \frac{1}{U_0''(0)^2} x
\eeas
as $t \searrow 0$, uniformly in $x \in S_1$. 

Since $X \subset C_S(B_3)$ there exists by Assertion 2 an element 
$\x \in C_S(B_3)$ such that
\[
\x - K\x = - q = - \frac{g}{U_0'} .
\]
This implies that $L_0 \x = g$ and thus that $L_0$ is onto, provided $\x \in X$.
To see the latter we observe that $\x = K\x + q$ is H\"older
continuous since $K\x$ is H\"older continuous. As above we conclude that
$V_\x \in C^2 (\R^3)$ and thus $K\x \in C^2(\dot B_3)$. Denoting by
$H_{V_\x}$ the Hessian of $V_\x$ we obtain for each $x \in \dot B_3$ a point
$z \in \overline{0,\, x}$ such that
\beas
\n{\nabla(K\x)(x)}
&\leq&
\left| \frac{U_0''(\n{x})}{U_0'(\n{x})^2}\right| \n{V_\x (x) - V_\x (0)}
+ \frac{1}{\n{U_0'(\n{x})}} \n{\nabla V_\x (x)} \\
&\leq&
\frac{C}{\n{x}^2} \left| \langle H_{V_\x} (z)x,x\rangle \right|
+ \frac{C}{\n{x}}  \n{\nabla V_\x (x)} 
\leq
C \nn{D^2 V_\x}_\infty .
\eeas
Finally, for $x \in S_1$ we have
\beas
\nabla (K\x) (t x)
&=&
- \frac{U_0''(t)}{U_0'(t)^2} x \bigl(V_\x (tx) - V_\x (0)\bigr) + \frac{1}{U_0'(t)}
\nabla V_\x (tx)\\
&=&
- U_0''(t)\left(\frac{t}{U_0'(t)}\right)^2  x \,\frac{1}{t^2} 
\, \frac{1}{2} \, \langle H_{V_\x} (\tau x)t x,t x\rangle +  \frac{t}{U_0'(t)}
\frac{\nabla V_\x (tx)}{t}\\
&\to&
- \frac{1}{2 U_0''(0)}\langle H_{V_\x} (0)x,x\rangle\, x + \frac{1}{U_0''(0)}
D^2 V_\x (0) x
\eeas
as $t \searrow 0$, uniformly in $x \in S_1$. We have shown that $K\x \in X$,
and since $q \in X$ as seen above this implies that $\x \in X$.
This completes the proof that $L_0$ is onto and thus also the proof of
Proposition~\ref{isom}. \prfe

\end{document}